\documentclass[aip,apl,amsmath,amssymb,reprint] {revtex4-1}

\usepackage{graphicx}% Include figure files
\usepackage{dcolumn}% Align table columns on decimal point
\usepackage{bm}% bold math
\usepackage{multirow}
\usepackage{xr}
\usepackage[dvipsnames]{xcolor}

\usepackage[utf8]{inputenc}
\usepackage[T1]{fontenc}
\usepackage{mathptmx}
\usepackage{etoolbox}

\makeatletter
\def\@email#1#2{%
 \endgroup
 \patchcmd{\titleblock@produce}
  {\frontmatter@RRAPformat}
  {\frontmatter@RRAPformat{\produce@RRAP{*#1\href{mailto:#2}{#2}}}\frontmatter@RRAPformat}
  {}{}
}%
\makeatother

\begin{document}

%\title{Electronic and Optical Properties of Bulk and Monolayer PdSe$_2$ from a Wannier-Localized Optimally-Tuned Screened Range-Separated Hybrid Functional}
\title{Resolving Contradictory Estimates of Band Gaps of Bulk PdSe$_2$: A Wannier-Localized Optimally-Tuned Screened Range-Separated Hybrid Density Functional Theory Study}

\author{Fred Florio}
    \affiliation{Department of Mechanical and Industrial Engineering, University of Massachusetts Amherst, Amherst MA 01003}
\author{María Camarasa-Gómez}
    \affiliation{Department of Molecular Chemistry and Materials Science, Weizmann Institute of Science, Rehovoth 7610001, Israel}
\author{Guy Ohad}
    \affiliation{Department of Molecular Chemistry and Materials Science, Weizmann Institute of Science, Rehovoth 7610001, Israel}
\author{Doron Naveh}
    \affiliation{Institute for Nanotechnology and Advanced Materials, Bar-Ilan University, Ramat Gan 52900, Israel}
\author{Leeor Kronik}
    \affiliation{Department of Molecular Chemistry and Materials Science, Weizmann Institute of Science, Rehovoth 7610001, Israel}
\author{Ashwin Ramasubramaniam}    
    \affiliation{Department of Mechanical and Industrial Engineering, University of Massachusetts Amherst, Amherst MA 01003}
    \affiliation{Materials Science and Engineering Graduate Program, University of Massachusetts Amherst, Amherst MA 01003}

\email{fflorio@umass.edu, ashwin@umass.edu}

\date{\today}

\begin{abstract}
Palladium diselenide (PdSe$_2$)---a layered van der Waals material---is attracting significant attention for optoelectronics due to the wide tunability of its band gap from the infrared through the visible range as a function of the number of layers. However, there continues to be disagreement over the precise nature and value of the optical band gap of bulk PdSe$_2$, owing to the rather small value of this gap that complicates experimental measurements and their interpretation. Here, we design and employ a  Wannier-localized optimally-tuned screened range-separated hybrid (WOT-SRSH) functional to investigate the electronic bandstructures and optical absorption spectra of bulk and monolayer PdSe$_2$. In particular, we account carefully for the finite exciton center-of-mass momentum within a time-dependent WOT-SRSH framework to calculate the \emph{indirect} optical gap and absorption onset accurately. Our results agree well with the best available photoconductivity measurements, as well as with state-of-the-art many-body perturbation theory calculations, confirming that bulk PdSe$_2$ has an optical gap in the mid-infrared (upper-bound of 0.44 eV). More generally, this work further bolsters the utility of the WOT-SRSH approach for predictive modeling of layered semiconductors.  
\end{abstract}

\maketitle

Layered van der Waals (vdW) materials have attracted significant attention over the last two decades, both in fundamental and applied research, due to their wide range of electronic, magnetic, optical, and structural behaviors.\cite{Geim2013,Novoselov2016,Ajayan2016,Review_LayerVdW_uses}
In particular, the ability of these vdW materials to be thinned down to the single- or few-layer limit has enabled wide-ranging discoveries in low-dimensional (2D) physics with the promise of novel applications in nanoscale optoelectronics.\cite{Xia2014,Review_LayerVdW_methods,Review_LayerVdW_stacking} 

Palladium diselenide (PdSe$_2$), a layered noble transition-metal dichalcogenide, is a particularly interesting member of the vdW family whose optical band gap is widely tunable over an approximate range of 0.5-1.5 eV when going from the bulk to the monolayer.\cite{ExpDFT_GapByLayer,ExpDFT_GapByLayer_Structure_Abs,Exp_GapByLayer_temp_IR}
The small bulk band gap, accompanied by its wide tunability with sample thickness, makes PdSe$_2$  particularly attractive for optoelectronics in the mid- to near-infrared (IR),\cite{Exp_IR_Stability,IRex1,IRex2} an application domain that has been explored more thoroughly with black phosphorus, which is unfortunately prone to degradation under normal ambient conditions.\cite{Island2015,Artel2017,Review_bP}
Indeed, recent work has demonstrated highly-responsive PdSe$_2$ IR photodetectors,\cite{Exp_GapByLayer_temp_IR,Exp_IR_Stability,IRex1,IRex2} as well as broadband devices,\cite{BBex1,Exp2_Abs_BBex2,BBex3,BBex4} that are stable during fabrication and operation.
Few-layer PdSe$_2$ is also known to exhibit unusual switching in its anisotropic linear dichroism response,\cite{Dichroism} providing an extra degree of control for optical switching, communication and  polarization-dependent detection.\cite{Review_PdSe2,Exp_IR_Stability}
Beyond optical response, ultra-thin field-effect transistors based on few-layer PdSe$_2$ have been shown to exhibit ambipolar behavior accompanied by high carrier mobilities.\cite{Ambipolar_mobility}
These examples, among others, highlight the promise of PdSe$_2$ and motivate continuing interest in this material.

Notwithstanding numerous investigations of PdSe$_2$, a definitive determination of its bulk band gap remains elusive.
An ARPES study\cite{Exp_ARPES} showed that bulk PdSe$_2$ is semiconducting, although the fundamental gap was not determined in that work.
Measurements of the optical band gap of bulk PdSe$_2$ range from 0.0-0.5eV\cite{ExpDFT_GapByLayer_Structure_Abs,Exp_GapByLayer_temp_IR,Exp2_Abs_BBex2,OldEXP} introducing uncertainty even in the qualitative behavior of this material (semi-metallic vs. semiconducting).
Nishiyama \emph{et al.}\,\cite{Nishiyama5} suggest that older absorption measurements\cite{ExpDFT_GapByLayer_Structure_Abs,Exp2_Abs_BBex2} are less reliable due to the large uncertainty introduced when determining optical gaps from Tauc plots. Instead, they estimated an indirect band gap of 0.5 eV (at 40 K) from photocurrent measurements that is consistent with photoresponse measurements by Zhang \emph{et al.}\cite{Exp_GapByLayer_temp_IR} 
On the computational front, density functional theory (DFT) calculations with semilocal functionals suggest the bulk is (nearly) metallic, with gaps ranging from 0-0.03 eV \cite{ExpDFT_GapByLayer_Structure_Abs,mBJ_Seebeck,Kim_Choi_GWgap} in clear disagreement with experiments.
A recent many-body perturbation theory (MBPT) study\cite{Kim_Choi_GWgap} based on the ``single-shot'' G$_0$W$_0$ approximation,\cite{GW_Louie} starting from the semilocal PBE functional, \cite{PBE} reported an indirect \emph{fundamental} gap of 0.45 eV.
We are unaware of any state-of-the-art calculations of the \emph{optical} gap of bulk PdSe$_2$, that typically employ the Bethe-Salpeter equation (BSE)\cite{BSE} to obtain neutral excitation energies starting from GW quasiparticle energies. Also, such GW-BSE calculations are usually performed only for direct optical transitions (zero-momentum excitons) and there are but a few examples of GW-BSE calculations for indirect transitions (finite-momentum excitons) in layered materials,\cite{FinMom1,FinMom2,FinMom3} a key requirement for determining the absorption onset of bulk PdSe$_2$ accurately.
In contrast to bulk PdSe$_2$, the situation for the monolayer is clearer, with experiment and theory agreeing, at least, on its semiconducting nature.
Experimental measurements of the optical band gap range between 1.25-1.37eV, \cite{ExpDFT_GapByLayer_Structure_Abs,Exp_GapByLayer_temp_IR} DFT calculations with semilocal functionals yield band gaps between 1.30-1.43 eV, \cite{ExpDFT_GapByLayer_Structure_Abs,Kuklin_Agren,mBJ_Seebeck} and a GW$_0$ study\cite{Kuklin_Agren} reported a quasiparticle gap of 2.55 eV.

While MBPT-based approaches are considered to be the state-of-the-art for first-principles electronic structure calculations, their accuracy comes at a high computational cost. Hence, there remains significant interest in developing less expensive yet equally accurate alternatives based on DFT and time-dependent DFT.
A promising approach along these lines is based on tuned screened range-separated hybrid (SRSH) exchange-correlation functionals that have been shown to deliver quantitatively accurate predictions of fundamental and optical band gaps in the solid state.\cite{SRSH1_alpBetEps2,SRSH2,Refaely-Abramson_15,Maria_WOT2}
More recently, these SRSH functionals have been constructed via a self-consistent framework, based on enforcing an ionization-potential \emph{ansatz} for maximally-localized Wannier functions\cite{Wannier90} to yield so-called Wannier-localized optimally-tuned SRSH (WOT-SRSH) functionals that are both quantitatively accurate in themselves and excellent starting points for subsequent MBPT calculations.\cite{WOT1,WOT2,WOT3,WOT4,Sagredo,Ohad_JChem_24}
Very recently, we have extended the WOT-SRSH framework to bulk and monolayer phases of black phosphorus (BP), molybdenum disulfide (MoS$_2$), and hexagonal boron nitride (hBN) and obtained accurate results.\cite{Maria_npj}
Here, we apply the WOT-SRSH approach to model electronic bandstructures and absorption spectra of bulk and monolayer PdSe$_2$. We show that the predictions are in good agreement with prior GW calculations,\cite{Kim_Choi_GWgap,Kuklin_Agren} as well as our own reference G$_0$W$_0$-BSE calculations.
We also calculate optical absorption spectra---accounting for both direct ($q \rightarrow 0$) and finite-momentum ($q\ne 0$) excitations---from which we find that the indirect optical gap of bulk PdSe$_2$ is 0.44 eV, in good agreement with the above mentioned photocurrent measurements.
Details of these calculations can be found in Sections S1 \& S2 of the Supporting Materials.

The SRSH functional is constructed by partitioning the $1/r$ Coulomb kernel as\cite{Coulomb1,Coulomb2}   
\begin{equation}
\frac{1}{r} =\frac{\alpha+\beta \text{erf}(\gamma r)}{r} +\frac{1-\left[\alpha+\beta \text{erf}(\gamma r)\right]}{r}
\label{eq:Coulomb}
\end{equation}
and treating the first and second terms of this equation using exact (Fock) exchange (XX) and semilocal (SL) DFT exchange, respectively; $\alpha, \beta, \gamma$ are parameters and $\text{erf}(\cdot)$ is the error function.
This procedure partitions the exchange interaction into short-range (SR) and long-range (LR) components with XX fractions of $\alpha$ and $\alpha+\beta$, respectively, accompanied by their respective complementary SL fractions of $1-\alpha$ and $1-(\alpha+\beta)$.
The range-separation parameter, $\gamma$, governs the transition from SR to LR behavior.
To enforce correct asymptotic screening of the Coulomb tail in the solid state, we set $\alpha+\beta=1/\epsilon_{\infty}$,\cite{Refaely-Abramson_13} where $\epsilon_{\infty}$ is the average high-frequency macroscopic dielectric constant (neglecting anisotropy for now).
The corresponding exchange potential, $v_{x}^{\text{SRSH}}$, is given by
\begin{equation}
v_{x}^{\text{SRSH}}  = \alpha v_{\text{XX}}^{\text{SR}}+(1-\alpha) v_{\text{SL}}^{\text{SR}}+\frac{v_{\text{XX}}^{\text{LR}}}{\epsilon_{\infty}}+\left(1-\frac{1}{\epsilon_{\infty}}\right)v_{\text{SL}}^{\text{LR}}.
\label{eq:potential}
\end{equation}
For monolayers, we set $\epsilon_{\infty}=1$, this being the exact asymptotic limit for screening in 2D;\cite{DieScreen1,DieScreen2,Qiu_etal} for bulk solids, $\epsilon_{\infty}$ is calculated \emph{ab initio} as $\epsilon_{\infty}=\rm{Tr}[\bm{\epsilon}_{\infty}]/3$.
To determine  $\alpha$ and $\gamma$, we enforce the ionization-potential (IP) \emph{ansatz} for removal of one electron from the highest-energy occupied, maximally-localized Wannier function ($\phi$).
In particular, an optimal $\alpha$-$\gamma$ pair must satisfy the relation
\begin{equation}
 E^{\alpha,\gamma}_{\text{constr}}[\phi](N-1)-E^{\alpha,\gamma}(N) = -\langle\phi | \hat{H}^{\alpha,\gamma}_{\text{SRSH}} | \phi \rangle,
 \label{eq:IPtheorem}
 \end{equation}
 where $E^{\alpha,\gamma}_{\text{constr}}[\phi](N-1)$ is the total energy of the system with an electron removed from $\phi$ (calculated using constrained minimization\cite{WOT1}), $E^{\alpha,\gamma}(N)$ is the total energy of the neutral system, and $\langle\phi | \hat{H}^{\alpha,\gamma}_\text{SRSH} | \phi\rangle$ is the expectation energy of $\phi$ with respect to the SRSH Hamiltonian.
 We direct the reader to Ref.\,\onlinecite{Maria_npj} for details of this procedure specialized to layered materials. 

\begin{figure}[h!]
\includegraphics[scale=0.5, angle=180, trim={0cm 0cm 2.2cm 1cm}]{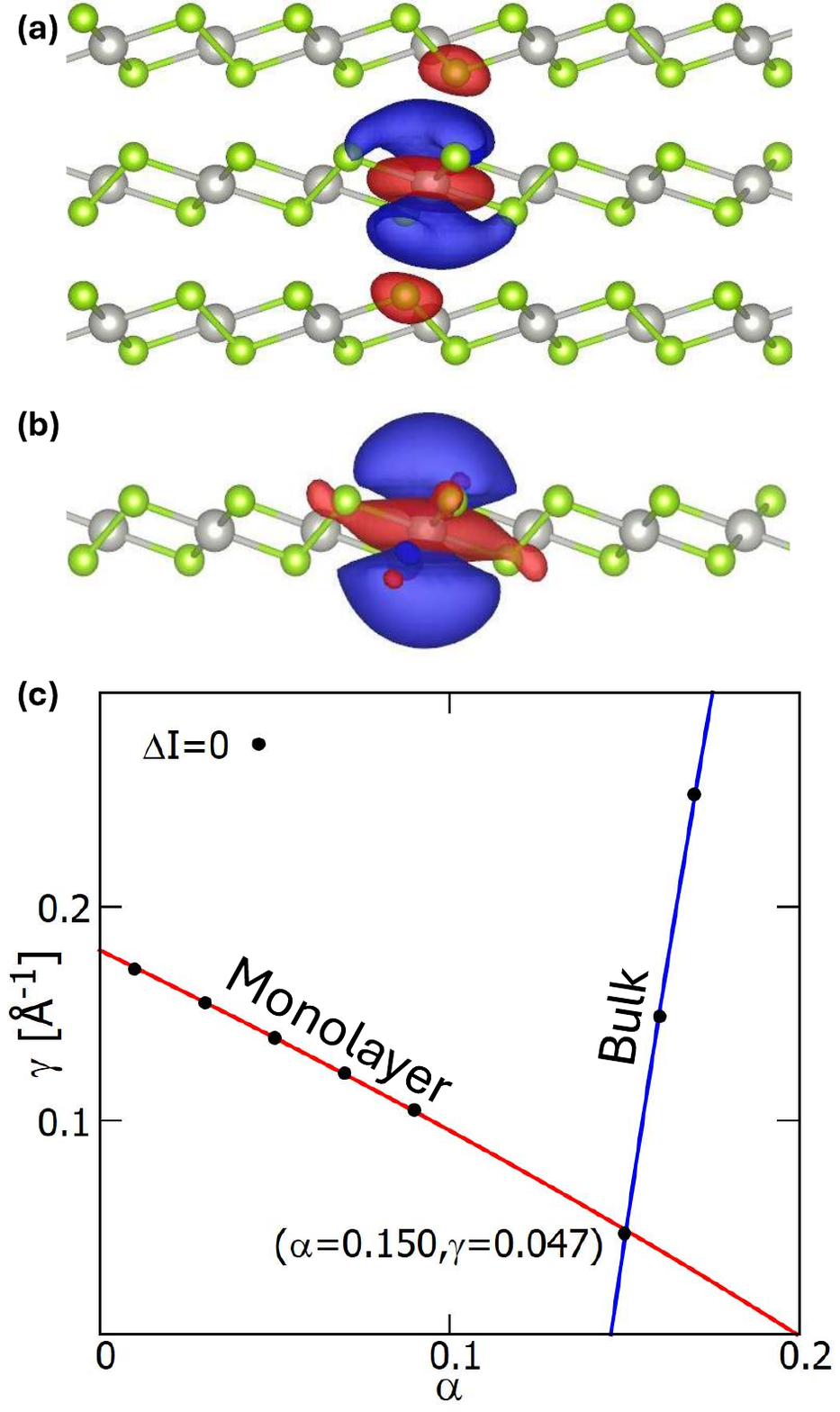}
\caption{\label{fig:WOT} Highest-energy occupied, maximally-localized Wannier function for (a) bulk and (b) monolayer PdSe$_2$.
(c) Curves satisfying the ionization potential (IP) \emph{ansatz}, $\Delta I \equiv  E^{\alpha,\gamma}_{constr}[\phi](N-1)-E^{\alpha,\gamma}(N)+\langle\phi | \hat{H}^{\alpha,\gamma}_{SRSH} | \phi \rangle$=0, for monolayer and bulk PdSe$_2$; their point of intersection, ($\alpha$ = 0.15 , $\gamma$ = 0.047\r{A}$^{-1}$), is optimal for both phases.}
\end{figure}

The outcome of the WOT procedure, along with plots of the Wannier functions employed for monolayer and bulk PdSe$_2$, are displayed in Figure\,\ref{fig:WOT}.
As shown previously,\cite{SRSH2,Maria_npj,Maria_WOT2} two reference structures are needed to determine $\alpha$ and $\gamma$ uniquely for vdW layered materials, the monolayer and bulk phases being particularly convenient choices.
While an infinite number of $\alpha$-$\gamma$ pairs satisfy the IP \textit{ansatz} (Eq.\,\ref{eq:IPtheorem}) for each phase, the intersection of the loci of these points for each phase results in a unique pair, $\alpha=0.15$ and $\gamma=0.047$ \AA$^{-1}$, that is simultaneously optimal for bulk and monolayer PdSe$_2$ (see Section S3 for the full set of tuning calculations). With this optimally-tuned functional now at hand, we proceed to predictive calculations of electronic bandstructure and absorption spectra. 

\begin{figure*}[htb!]
\includegraphics[angle=270, width=\textwidth, trim={3cm 1cm 6cm 1cm}]{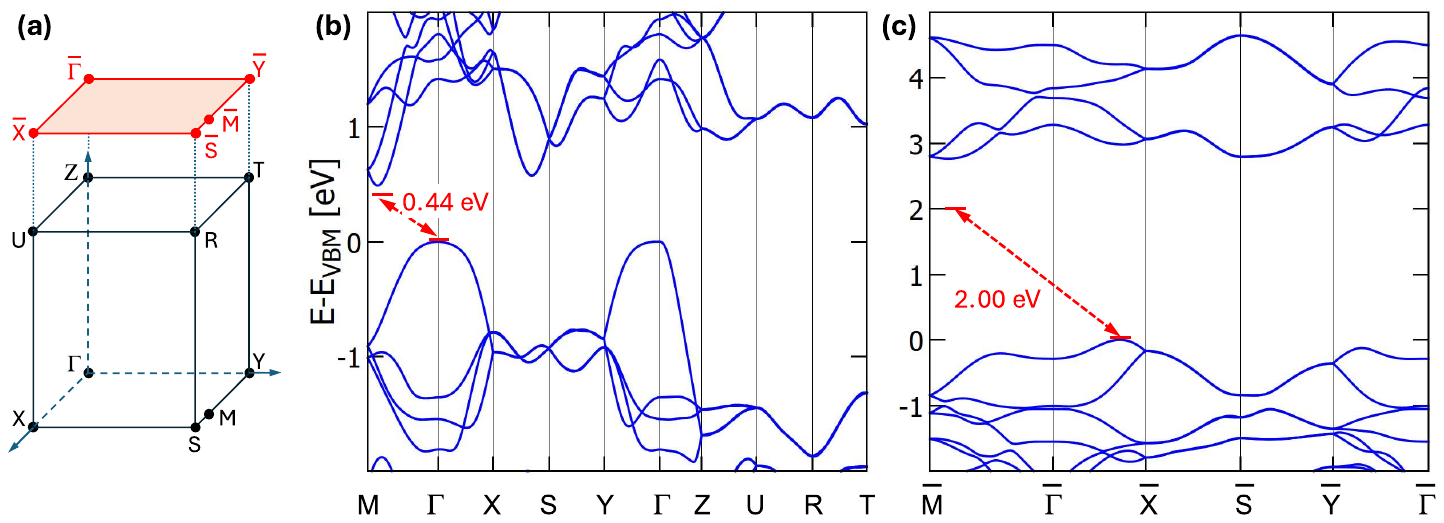}
\caption{\label{fig:Ebands}
(a) Schematic of reduced Brillouin zones for bulk (black) and monolayer (red) PdSe$_2$.
Wannier-interpolated\cite{Wannier90} bandstructures for (b) bulk and (c) monolayer PdSe$_2$ calculated using the WOT-SRSH functional with $12\times 12 \times 9$ and $24\times 24 \times 1$ $k$-point meshes.
For bulk PdSe$_2$, the valence band maximum is located at the $\Gamma$ point while the conduction band minimum is located in the interior of the Brillouin zone along the $\Gamma-\rm{M}$ direction where M=(0.40, 0.50, 0.00) in reciprocal coordinates.
For the monolayer, the valence band maximum is located along the $\bar{\Gamma}-\bar{\rm X}$ direction while the conduction band minimum is located in the interior of the Brillouin zone along the $\bar{\Gamma}-\bar{\rm{M}}$ direction where $\bar{\rm{M}}$=(0.42, 0.50, 0.00) in reciprocal coordinates.
Indirect optical band gaps are also indicated schematically in (b, c) by the red arrows; all fundamental and optical band gaps are reported in Table \ref{table:SRSH}. 
}
\end{figure*}

Figure \ref{fig:Ebands} displays WOT-SRSH bandstructures for bulk and monolayer PdSe$_2$,  using well-converged $k$-point meshes; direct and indirect band-gaps are reported in Table \ref{table:SRSH}.
Both bulk and monolayer PdSe$_2$ are indirect gap semiconductors and our calculations further confirm that the conduction band minima lie in the interior of the Brillouin zone (along the $\Gamma-\rm{M}$/$\bar{\Gamma}-\bar{\rm{M}}$ line) rather than along a high-symmetry direction.\cite{Kim_Choi_GWgap, Kuklin_Agren}
The indirect bulk band gap is calculated to be 0.49 eV, which agrees well with Kim and Choi's estimate of 0.45 eV.\cite{Kim_Choi_GWgap}
The indirect gap for the monolayer is calculated to be 2.76 eV, which is slightly higher than  Kim and Choi's estimate of 2.37 eV\cite{Kim_Choi_GWgap} but closer to Kuklin and {\AA}gren's estimate of 2.55 eV.\cite{Kuklin_Agren}
For comparison, we also calculated quasiparticle band gaps independently by performing single-shot GW calculations with HSE06\cite{HSE06} wavefunctions as the starting point (Section S5).
Using less dense $8\times 8 \times 6$ $k$-point meshes---necessitated by the high computational cost of full-frequency GW--- for both the GW and WOT-SRSH calculations, the bulk band gap is estimated to be 0.6 eV and 0.57 eV, respectively. As seen in our prior GW calculations,\cite{SRSH2,Maria_WOT2} converging the calculation with respect to the $k$-point mesh is expected to reduce the GW gap and bring it closer to the the converged WOT-SRSH estimate.
Similarly, for the monolayer, a comparison of GW and WOT-SRSH calculations using $14\times 14\times 1$ $k$-point meshes reveals very similar indirect gaps of 2.73 eV and 2.76 eV, respectively.
It is well known that GW calculations are sensitive to the choice of starting point; the WOT-SRSH approach, on the other hand is not only internally self-consistent but has also been shown to be a good starting point for single-shot GW calculations.\cite{WOT3,Maria_npj}
This fact, bolstered by the good agreement between our WOT-SRSH and single-shot GW (over HSE) results, leads us to believe that our calculated fundamental band gaps are reliable.
While we are unaware of any direct measurements of the fundamental band gap of bulk PdSe$_2$, electrical transport  gaps range from 0.3 eV (18-191 nm thick flakes) \cite{Nishiyama3} to 0.57 eV (6.8 nm thick flake),\cite{Exp_GapByLayer_temp_IR} but these results could be affected by the presence of trap states as well inherent uncertainties in device modeling parameters (see Section S4 for a more complete comparison of calculated and measured band gaps).
For completeness, we note that inclusion of spin-orbit coupling alters our estimates of both bulk and monolayer WOT-SRSH band gaps by $\sim$10 meV with insignificant differences in the overall band structure (Section S5). 

\begin{table*}[]
\caption{\label{table:SRSH}%
WOT-SRSH parameters ($\alpha$, $\gamma$), average high-frequency macroscopic dielectric constant ($\epsilon^{\infty}$), Brillouin zone sampling, direct/indirect fundamental band gaps (E$^{f}_{d/i}$), and direct/indirect optical band gaps (E$^{opt}_{d/i}$) for monolayer and bulk PdSe$_2$; literature data from GW and GW-BSE calculations are provided where available.
}
\begin{ruledtabular}
\begin{tabular}{ccccccccc}
\textrm{Phase}&
\textrm{$\alpha$}&
\textrm{$\gamma$ [\r{A}$^{-1}$]}&
\textrm{$\epsilon_{\infty}$}&
\textrm{$\bm{k}$-grid}&
\textrm{E$^{f}_{i}$ [eV]}&
\textrm{E$^{f}_{d}$ [eV]}&
\textrm{E$^{opt}_{i}$ [eV]}&
\textrm{E$^{opt}_{d}$ [eV]}\\

\colrule

Bulk & 0.150 & 0.047 & 12.136 & $12\times 12\times 9$ & 0.49 (0.45 \cite{Kim_Choi_GWgap}) & 1.27 (1.10 \cite{Kim_Choi_GWgap}) & 0.44 & 1.11 \\ 
Monolayer & 0.150 & 0.047 & 1 & $24\times 24\times 1$ & 2.76 (2.37,\cite{Kim_Choi_GWgap} 2.53 \cite{Kuklin_Agren})& 2.98 (2.54,\cite{Kim_Choi_GWgap} 2.62\cite{Kuklin_Agren} ) & 2.00 & 2.20 (2.01\cite{Kuklin_Agren}) \\
\end{tabular}
\end{ruledtabular}
\end{table*}

\begin{figure*}[]
\includegraphics[width=8cm, angle=270, trim={3cm 0cm 6cm 0cm}]{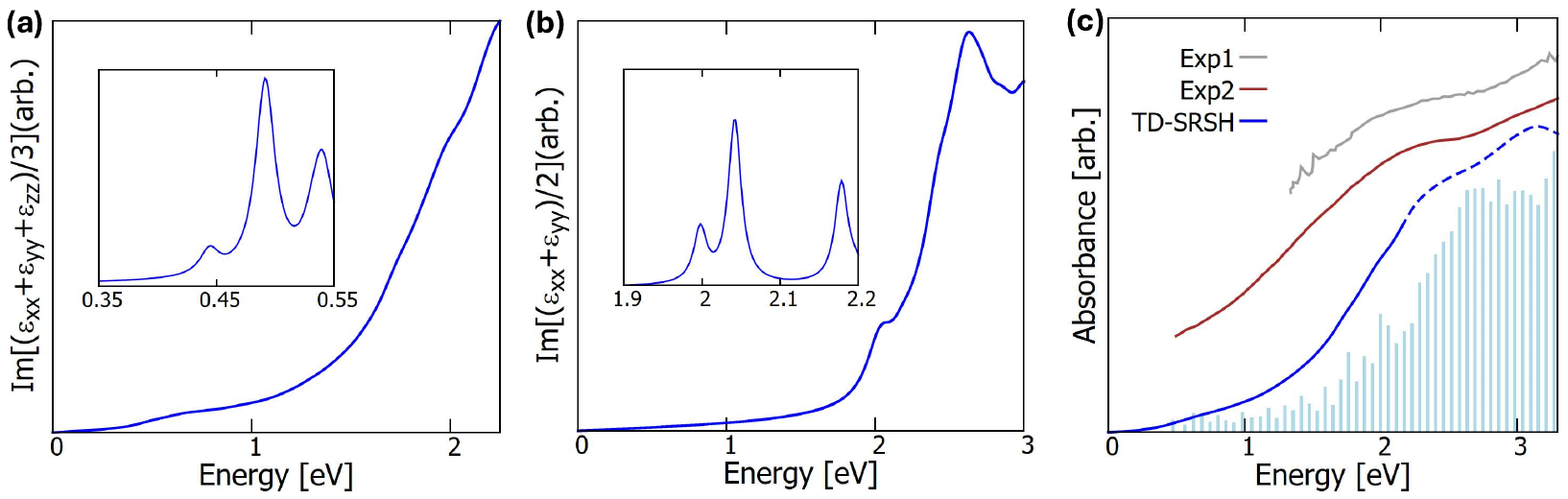}
\caption{\label{fig:AbsSpectra}
TD-SRSH absorption spectra of (a) bulk and (b) monolayer PdSe$_2$ for finite momentum (${\bm q} \to {\bm Q}$) excitons with spectral broadening of 0.1 eV.
The insets in (a) and (b) show a smaller energy window near the absorption onset with smaller spectral broadening (0.01 eV), resolving the optical transitions more clearly (see Table \ref{table:SRSH} for numerical values).
(c) Tauc plots displaying the absorbance for bulk PdSe$_2$ calculated from the ${\bm q} \to {\bm Q}$ spectrum (0.1 eV broadening) along with experimental data from the literature [Exp1: 42-layer ($\sim$16 nm) flakes; \cite{ExpDFT_GapByLayer_Structure_Abs} Exp2: 50-layer ($\sim$20 nm) flakes \cite{Exp2_Abs_BBex2}]. The curves have been shifted vertically for clarity. Vertical bars represent cumulative oscillator strengths at various energies. The calculations capture all optical transitions up to $2$ eV; the dashed portion of the calculated curve signifies that the absorption spectrum above 2 eV is incomplete. The absorbance is calculated as $\sqrt{\alpha(\omega) h\nu}$ where $\alpha(\omega)=2\omega\kappa(\omega)/c$, $\kappa(\omega)$ being the extinction coefficient.\cite{grosso2000solid} 
}
\end{figure*}

To make a clear connection with optical measurements, we calculated the optical response of bulk and monolayer PdSe$_2$ using linear-response, time-dependent DFT calculations that employ the WOT-SRSH functional (TD-SRSH).
We have used this approach successfully to model the absorption spectra and to extract exciton binding energies for several 2D materials, albeit only for vertical transitions, i.e., when excitons couple to incoming photons with vanishingly small momenta (${\bm q} \to 0)$.\cite{SRSH2,Maria_WOT2,Maria_npj} 
However, an exciton can indeed carry a finite center-of-mass momentum, ${\bm Q}$, and this is central to understanding exciton dispersion and excitonic bandstructures of materials;\cite{FinMom1,ExcitonDispersion,BeyondTamDan,FinMomExcitons1,Noffsinger} we refer the reader to the cited literature for details and focus on the practical aspects of this problem for now.
Specifically, our interest here is to employ finite-momentum calculations, within the framework of TD-SRSH, to calculate the \emph{indirect} optical band gaps (Fig.\,\ref{fig:Ebands}) and thereby resolve contradictory reports in the literature, in particular, for bulk PdSe$_2$. 
To this end, we assign finite exciton center-of-mass momenta ${\bm Q}=(0.3333, 0.4167,
 0.0000)$ and ${\bm Q}=(0.4167, 0.2917, 0.0000)$ to the bulk and monolayer, respectively, and calculate the excitation spectrum;\footnotemark[1] we do not include  electron-phonon coupling terms\cite{Noffsinger} in the optical matrix elements for now, deferring this to future work.
\footnotetext[1]{The center-of-mass momentum vector, ${\bm Q}$, is denoted here as a multiple of the reciprocal lattice vectors.
The band-edge extrema obtained from Wannier interpolation (Fig.\,\ref{fig:Ebands}) do not necessarily coincide with sampling points in the regular $k$-point mesh used for self-consistent calculations.
As VASP requires $\bm Q$ to be expressible as a difference between two $k$-points, not allowing for interpolation to irregular points, we pick $k$-points that are as close as possible to the valence band maximum and the conduction band minimum.
Thus, our estimates for optical gaps are upper bounds.
The resulting overestimate of the optical gap can be approximated by the difference of the electronic band gap between the two sampled points and the true indirect fundamental band gap, and we estimate this error to be 13.5 meV for the bulk and 5.3 meV for the monolayer.
The error can be reduced with denser $k$-point grids.
}
Figure \ref{fig:AbsSpectra}(a) displays the absorption spectra for transitions with ${\bm q} \to {\bm Q}$; spectra for ${\bm q} \to 0$ transitions are provided in Section S6. Resolving the absorption spectra more clearly (Fig.\,\ref{fig:AbsSpectra}(a), inset), we find that the lowest indirect optical transition (${\bm q} \to {\bm Q}$) for bulk PdSe$_2$ is at 0.44 eV, \footnotemark[1] approximately, 0.05 eV below the fundamental indirect gap.
The lowest direct optical transition (${\bm q} \to 0$) is at much higher energy (1.11 eV) and not directly relevant to the absorption onset, underscoring the need for finite-momentum calculations.
To reinforce the validity of the TD-SRSH results, we also calculated these optical transitions using the GW-BSE  approach (using a coarser $k$-grid, as noted above).
The two sets of calculations, based on entirely different methodologies, are in good agreement (Sections S6 \& S7), yielding indirect optical gaps of 0.58 eV (GW-BSE) and 0.47 eV (TD-SRSH). 
For the monolayer, we find TD-SRSH optical gaps of  2.00 eV (indirect) and 2.20 eV (direct), which implies that excitons are strongly bound ($\sim$0.75 eV binding energy) in this 2D system.\cite{AR2012}
Our TD-SRSH estimates for direct optical gaps are slightly higher but still in good agreement with other reports (2.01 eV; GW-BSE \cite{Kuklin_Agren}); both direct and indirect TD-SRSH gaps are also in good agreement (to within 100 meV) with our own GW-BSE results (Section S7).

Our results now provide us with a clearer understanding of experimental reports, specifically, absorption measurements.
As absorption data are typically presented as Tauc plots (absorbance vs.\,energy), from which band gaps are determined by linear extrapolation, we replot the bulk finite-momentum optical spectrum in this form in Figure\,\ref{fig:AbsSpectra}, alongside data from two experiments.\cite{ExpDFT_GapByLayer_Structure_Abs, Exp2_Abs_BBex2}
We intentionally use a relatively large broadening of 0.1 eV to mimic the experimental data and emphasize that the tail below 0.44 eV is simply an artifact of this broadening.\footnotemark[2]
\footnotetext[2]{Note that the actual absorption spectrum will contain contributions from multiple $\bm Q$-vectors but we are only concerned with one momentum here to capture the absorption onset.}
Oyedele \emph{et al.}'s absorption measurements (Exp1 in Fig.\,\ref{fig:AbsSpectra}) stop at $\sim$1.3 eV, requiring linear extrapolation over a significant energy window to determine the absorption onset.
Zeng \emph{et al.}'s measurements extend to lower energies of $\sim$0.5 eV, but the nonlinearity in the data introduces errors in the extrapolation.
These issues with the extrapolation procedure were also noted by Nishiyama \emph{et al.}\cite{Nishiyama3}
Conversely, band gaps inferred from photocurrent measurements ($\sim$0.5 eV \cite{Exp_GapByLayer_temp_IR,Nishiyama5}) are in excellent agreement with our calculations and, collectively, theory and experiment confirm that bulk PdSe$_2$ (in its pristine state) is indeed a small-gap semiconductor with absorption onset in the mid-infrared.

In conclusion, we have designed and employed a Wannier-optimally tuned screened range-separated (WOT-SRSH) hybrid functional  to model the optoelectronic properties of bulk and monolayer PdSe$_2$.
This methodology is internally self-consistent and nonempirical, and delivers results that are in quantitative agreement with state-of-the-art GW-BSE calculations as well as the best available photocurrent measurements of bulk PdSe$_2$.
Our calculation of the finite-momentum excitonic spectrum is key to achieving this agreement with experiments, given that PdSe$_2$ is an indirect gap material with a substantial difference between the direct and indirect gap.
The straightforward resolution of conflicting reports in the literature provided here further confirms the potential of WOT-SRSH functionals for predictive modeling of semiconductors and insulators.
 
 \vspace{0.1cm} 
See the supplementary material for details of electronic-structure calculations; atomic positions and $k$-point sampling; data from WOT tuning calculations; consolidated tables of electronic and optical band gaps from the literature and this work; and comparisons of GW/GW-BSE and WOT-SRSH/TD-SRSH results.

This work was supported by the National Science Foundation (NSF-BSF 2150562), the U.S.-Israel Binational Science Foundation (grant no. 2021721), and the Israel Science Foundation.
M.C.-G. is grateful to the Azrieli Foundation for the award of an Azrieli International Postdoctoral Fellowship.
L.K. acknowledges support from the Aryeh and Mintzi Katzman Professorial Chair, and the Helen and Martin Kimmel Award for Innovative Investigation.

\section*{Author Declarations}
\subsection*{Conflict of Interest}
The authors have no conflicts to disclose.

\subsection*{Author contributions}
{\bf Fred Florio:} Data curation (lead); Formal Analysis (equal); Methodology (equal); Visualization (lead); Writing -- original draft (equal); 
{\bf María Camarasa-Gómez:} Methodology (equal); Writing -- review \& editing (equal)
{\bf Guy Ohad:}  Methodology (equal); Writing -- review \& editing (equal)
{\bf Doron Naveh:} Conceptualization (equal); Funding Acquisition (equal); Writing -- review \& editing (equal)
{\bf Leeor Kronik:} Funding Acquisition (equal); Writing -- review \& editing (equal)
{\bf Ashwin Ramasubramaniam:} Conceptualization (lead); Formal analysis (equal); Funding Acquisition (equal); Methodology (equal); Supervision (lead); Visualization (equal); Writing -- original draft (equal)

\section*{Data Availability}
The data that support the findings of this study are available from the corresponding authors upon reasonable request.

\bibliography{refs}% Produces the bibliography via BibTeX.

\providecommand{\noopsort}[1]{}\providecommand{\singleletter}[1]{#1}%
\begin{thebibliography}{63}%
\makeatletter
\providecommand \@ifxundefined [1]{%
 \@ifx{#1\undefined}
}%
\providecommand \@ifnum [1]{%
 \ifnum #1\expandafter \@firstoftwo
 \else \expandafter \@secondoftwo
 \fi
}%
\providecommand \@ifx [1]{%
 \ifx #1\expandafter \@firstoftwo
 \else \expandafter \@secondoftwo
 \fi
}%
\providecommand \natexlab [1]{#1}%
\providecommand \enquote  [1]{``#1''}%
\providecommand \bibnamefont  [1]{#1}%
\providecommand \bibfnamefont [1]{#1}%
\providecommand \citenamefont [1]{#1}%
\providecommand \href@noop [0]{\@secondoftwo}%
\providecommand \href [0]{\begingroup \@sanitize@url \@href}%
\providecommand \@href[1]{\@@startlink{#1}\@@href}%
\providecommand \@@href[1]{\endgroup#1\@@endlink}%
\providecommand \@sanitize@url [0]{\catcode `\\12\catcode `\$12\catcode
  `\&12\catcode `\#12\catcode `\^12\catcode `\_12\catcode `\%12\relax}%
\providecommand \@@startlink[1]{}%
\providecommand \@@endlink[0]{}%
\providecommand \url  [0]{\begingroup\@sanitize@url \@url }%
\providecommand \@url [1]{\endgroup\@href {#1}{\urlprefix }}%
\providecommand \urlprefix  [0]{URL }%
\providecommand \Eprint [0]{\href }%
\providecommand \doibase [0]{http://dx.doi.org/}%
\providecommand \selectlanguage [0]{\@gobble}%
\providecommand \bibinfo  [0]{\@secondoftwo}%
\providecommand \bibfield  [0]{\@secondoftwo}%
\providecommand \translation [1]{[#1]}%
\providecommand \BibitemOpen [0]{}%
\providecommand \bibitemStop [0]{}%
\providecommand \bibitemNoStop [0]{.\EOS\space}%
\providecommand \EOS [0]{\spacefactor3000\relax}%
\providecommand \BibitemShut  [1]{\csname bibitem#1\endcsname}%
\let\auto@bib@innerbib\@empty
%</preamble>
\bibitem [{\citenamefont {Geim}\ and\ \citenamefont
  {Grigorieva}(2013)}]{Geim2013}%
  \BibitemOpen
  \bibfield  {author} {\bibinfo {author} {\bibfnamefont {A.}~\bibnamefont
  {Geim}}\ and\ \bibinfo {author} {\bibfnamefont {I.}~\bibnamefont
  {Grigorieva}},\ }\href@noop {} {\bibfield  {journal} {\bibinfo  {journal}
  {Nature}\ }\textbf {\bibinfo {volume} {499}},\ \bibinfo {pages} {419}
  (\bibinfo {year} {2013})}\BibitemShut {NoStop}%
\bibitem [{\citenamefont {Novoselov}\ \emph {et~al.}(2016)\citenamefont
  {Novoselov}, \citenamefont {Mishchenko}, \citenamefont {Carvalho},\ and\
  \citenamefont {Castro~Neto}}]{Novoselov2016}%
  \BibitemOpen
  \bibfield  {author} {\bibinfo {author} {\bibfnamefont {K.~S.}\ \bibnamefont
  {Novoselov}}, \bibinfo {author} {\bibfnamefont {A.}~\bibnamefont
  {Mishchenko}}, \bibinfo {author} {\bibfnamefont {A.}~\bibnamefont
  {Carvalho}}, \ and\ \bibinfo {author} {\bibfnamefont {A.~H.}\ \bibnamefont
  {Castro~Neto}},\ }\href@noop {} {\bibfield  {journal} {\bibinfo  {journal}
  {Science}\ }\textbf {\bibinfo {volume} {353}},\ \bibinfo {pages} {aac9439}
  (\bibinfo {year} {2016})}\BibitemShut {NoStop}%
\bibitem [{\citenamefont {Ajayan}, \citenamefont {Kim},\ and\ \citenamefont
  {Banerjee}(2016)}]{Ajayan2016}%
  \BibitemOpen
  \bibfield  {author} {\bibinfo {author} {\bibfnamefont {P.}~\bibnamefont
  {Ajayan}}, \bibinfo {author} {\bibfnamefont {P.}~\bibnamefont {Kim}}, \ and\
  \bibinfo {author} {\bibfnamefont {K.}~\bibnamefont {Banerjee}},\ }\href@noop
  {} {\bibfield  {journal} {\bibinfo  {journal} {Phys. Today}\ }\textbf
  {\bibinfo {volume} {69}},\ \bibinfo {pages} {38} (\bibinfo {year}
  {2016})}\BibitemShut {NoStop}%
\bibitem [{\citenamefont {Duong}, \citenamefont {Yun},\ and\ \citenamefont
  {Lee}(2017)}]{Review_LayerVdW_uses}%
  \BibitemOpen
  \bibfield  {author} {\bibinfo {author} {\bibfnamefont {D.~L.}\ \bibnamefont
  {Duong}}, \bibinfo {author} {\bibfnamefont {S.~J.}\ \bibnamefont {Yun}}, \
  and\ \bibinfo {author} {\bibfnamefont {Y.~H.}\ \bibnamefont {Lee}},\
  }\href@noop {} {\bibfield  {journal} {\bibinfo  {journal} {ACS Nano}\
  }\textbf {\bibinfo {volume} {11}},\ \bibinfo {pages} {11803} (\bibinfo {year}
  {2017})}\BibitemShut {NoStop}%
\bibitem [{\citenamefont {Xia}\ \emph {et~al.}(2014)\citenamefont {Xia},
  \citenamefont {Wang}, \citenamefont {Xiao}, \citenamefont {Dubey},\ and\
  \citenamefont {Ramasubramaniam}}]{Xia2014}%
  \BibitemOpen
  \bibfield  {author} {\bibinfo {author} {\bibfnamefont {F.}~\bibnamefont
  {Xia}}, \bibinfo {author} {\bibfnamefont {H.}~\bibnamefont {Wang}}, \bibinfo
  {author} {\bibfnamefont {D.}~\bibnamefont {Xiao}}, \bibinfo {author}
  {\bibfnamefont {M.}~\bibnamefont {Dubey}}, \ and\ \bibinfo {author}
  {\bibfnamefont {A.}~\bibnamefont {Ramasubramaniam}},\ }\href@noop {}
  {\bibfield  {journal} {\bibinfo  {journal} {Nature Photonics}\ }\textbf
  {\bibinfo {volume} {8}},\ \bibinfo {pages} {899} (\bibinfo {year}
  {2014})}\BibitemShut {NoStop}%
\bibitem [{\citenamefont {Zhou}\ \emph {et~al.}(2018)\citenamefont {Zhou},
  \citenamefont {Hu}, \citenamefont {Yu}, \citenamefont {Liu}, \citenamefont
  {Shu}, \citenamefont {Zhang}, \citenamefont {Li}, \citenamefont {Ma},
  \citenamefont {Xu},\ and\ \citenamefont {Zhai}}]{Review_LayerVdW_methods}%
  \BibitemOpen
  \bibfield  {author} {\bibinfo {author} {\bibfnamefont {X.}~\bibnamefont
  {Zhou}}, \bibinfo {author} {\bibfnamefont {X.}~\bibnamefont {Hu}}, \bibinfo
  {author} {\bibfnamefont {J.}~\bibnamefont {Yu}}, \bibinfo {author}
  {\bibfnamefont {S.}~\bibnamefont {Liu}}, \bibinfo {author} {\bibfnamefont
  {Z.}~\bibnamefont {Shu}}, \bibinfo {author} {\bibfnamefont {Q.}~\bibnamefont
  {Zhang}}, \bibinfo {author} {\bibfnamefont {H.}~\bibnamefont {Li}}, \bibinfo
  {author} {\bibfnamefont {Y.}~\bibnamefont {Ma}}, \bibinfo {author}
  {\bibfnamefont {H.}~\bibnamefont {Xu}}, \ and\ \bibinfo {author}
  {\bibfnamefont {T.}~\bibnamefont {Zhai}},\ }\href@noop {} {\bibfield
  {journal} {\bibinfo  {journal} {Adv. Functional Mater.}\ }\textbf {\bibinfo
  {volume} {28}},\ \bibinfo {pages} {1706587} (\bibinfo {year}
  {2018})}\BibitemShut {NoStop}%
\bibitem [{\citenamefont {Zhang}\ \emph {et~al.}(2016)\citenamefont {Zhang},
  \citenamefont {Wang}, \citenamefont {Chen}, \citenamefont {Wang},\ and\
  \citenamefont {Wee}}]{Review_LayerVdW_stacking}%
  \BibitemOpen
  \bibfield  {author} {\bibinfo {author} {\bibfnamefont {W.}~\bibnamefont
  {Zhang}}, \bibinfo {author} {\bibfnamefont {Q.}~\bibnamefont {Wang}},
  \bibinfo {author} {\bibfnamefont {Y.}~\bibnamefont {Chen}}, \bibinfo {author}
  {\bibfnamefont {Z.}~\bibnamefont {Wang}}, \ and\ \bibinfo {author}
  {\bibfnamefont {A.~T.~S.}\ \bibnamefont {Wee}},\ }\href@noop {} {\bibfield
  {journal} {\bibinfo  {journal} {2D Mater.}\ }\textbf {\bibinfo {volume}
  {3}},\ \bibinfo {pages} {022001} (\bibinfo {year} {2016})}\BibitemShut
  {NoStop}%
\bibitem [{\citenamefont {Wei}\ \emph {et~al.}(2022)\citenamefont {Wei},
  \citenamefont {Lian}, \citenamefont {Zhang}, \citenamefont {Wang},
  \citenamefont {Wang},\ and\ \citenamefont {Xu}}]{ExpDFT_GapByLayer}%
  \BibitemOpen
  \bibfield  {author} {\bibinfo {author} {\bibfnamefont {M.}~\bibnamefont
  {Wei}}, \bibinfo {author} {\bibfnamefont {J.}~\bibnamefont {Lian}}, \bibinfo
  {author} {\bibfnamefont {Y.}~\bibnamefont {Zhang}}, \bibinfo {author}
  {\bibfnamefont {C.}~\bibnamefont {Wang}}, \bibinfo {author} {\bibfnamefont
  {Y.}~\bibnamefont {Wang}}, \ and\ \bibinfo {author} {\bibfnamefont
  {Z.}~\bibnamefont {Xu}},\ }\href@noop {} {\bibfield  {journal} {\bibinfo
  {journal} {npj 2D Mater. and Appl.}\ }\textbf {\bibinfo {volume} {6}},\
  \bibinfo {pages} {1} (\bibinfo {year} {2022})}\BibitemShut {NoStop}%
\bibitem [{\citenamefont {Oyedele}\ \emph {et~al.}(2017)\citenamefont
  {Oyedele}, \citenamefont {Yang}, \citenamefont {Liang}, \citenamefont
  {Puretzky}, \citenamefont {Wang}, \citenamefont {Zhang}, \citenamefont {Yu},
  \citenamefont {Pudasaini}, \citenamefont {Ghosh}, \citenamefont {Liu},
  \citenamefont {Rouleau}, \citenamefont {Sumpter}, \citenamefont {Chisholm},
  \citenamefont {Zhou}, \citenamefont {Rack}, \citenamefont {Geohegan},\ and\
  \citenamefont {Xiao}}]{ExpDFT_GapByLayer_Structure_Abs}%
  \BibitemOpen
  \bibfield  {author} {\bibinfo {author} {\bibfnamefont {A.~D.}\ \bibnamefont
  {Oyedele}}, \bibinfo {author} {\bibfnamefont {S.}~\bibnamefont {Yang}},
  \bibinfo {author} {\bibfnamefont {L.}~\bibnamefont {Liang}}, \bibinfo
  {author} {\bibfnamefont {A.~A.}\ \bibnamefont {Puretzky}}, \bibinfo {author}
  {\bibfnamefont {K.}~\bibnamefont {Wang}}, \bibinfo {author} {\bibfnamefont
  {J.}~\bibnamefont {Zhang}}, \bibinfo {author} {\bibfnamefont
  {P.}~\bibnamefont {Yu}}, \bibinfo {author} {\bibfnamefont {P.~R.}\
  \bibnamefont {Pudasaini}}, \bibinfo {author} {\bibfnamefont {A.~W.}\
  \bibnamefont {Ghosh}}, \bibinfo {author} {\bibfnamefont {Z.}~\bibnamefont
  {Liu}}, \bibinfo {author} {\bibfnamefont {C.~M.}\ \bibnamefont {Rouleau}},
  \bibinfo {author} {\bibfnamefont {B.~G.}\ \bibnamefont {Sumpter}}, \bibinfo
  {author} {\bibfnamefont {M.~F.}\ \bibnamefont {Chisholm}}, \bibinfo {author}
  {\bibfnamefont {W.}~\bibnamefont {Zhou}}, \bibinfo {author} {\bibfnamefont
  {P.~D.}\ \bibnamefont {Rack}}, \bibinfo {author} {\bibfnamefont {D.~B.}\
  \bibnamefont {Geohegan}}, \ and\ \bibinfo {author} {\bibfnamefont
  {K.}~\bibnamefont {Xiao}},\ }\href@noop {} {\bibfield  {journal} {\bibinfo
  {journal} {J. of the Am. Chem. Soc.}\ }\textbf {\bibinfo {volume} {139}},\
  \bibinfo {pages} {14090} (\bibinfo {year} {2017})}\BibitemShut {NoStop}%
\bibitem [{\citenamefont {Zhang}\ \emph {et~al.}(2019)\citenamefont {Zhang},
  \citenamefont {Amani}, \citenamefont {Chaturvedi}, \citenamefont {Tan},
  \citenamefont {Bullock}, \citenamefont {Song}, \citenamefont {Kim},
  \citenamefont {Lien}, \citenamefont {Scott}, \citenamefont {Zhang},\ and\
  \citenamefont {Javey}}]{Exp_GapByLayer_temp_IR}%
  \BibitemOpen
  \bibfield  {author} {\bibinfo {author} {\bibfnamefont {G.}~\bibnamefont
  {Zhang}}, \bibinfo {author} {\bibfnamefont {M.}~\bibnamefont {Amani}},
  \bibinfo {author} {\bibfnamefont {A.}~\bibnamefont {Chaturvedi}}, \bibinfo
  {author} {\bibfnamefont {C.}~\bibnamefont {Tan}}, \bibinfo {author}
  {\bibfnamefont {J.}~\bibnamefont {Bullock}}, \bibinfo {author} {\bibfnamefont
  {X.}~\bibnamefont {Song}}, \bibinfo {author} {\bibfnamefont {H.}~\bibnamefont
  {Kim}}, \bibinfo {author} {\bibfnamefont {D.-H.}\ \bibnamefont {Lien}},
  \bibinfo {author} {\bibfnamefont {M.~C.}\ \bibnamefont {Scott}}, \bibinfo
  {author} {\bibfnamefont {H.}~\bibnamefont {Zhang}}, \ and\ \bibinfo {author}
  {\bibfnamefont {A.}~\bibnamefont {Javey}},\ }\href@noop {} {\bibfield
  {journal} {\bibinfo  {journal} {Appl. Phys. Lett.}\ }\textbf {\bibinfo
  {volume} {114}},\ \bibinfo {pages} {253102} (\bibinfo {year}
  {2019})}\BibitemShut {NoStop}%
\bibitem [{\citenamefont {Long}\ \emph {et~al.}(2019)\citenamefont {Long},
  \citenamefont {Wang}, \citenamefont {Wang}, \citenamefont {Zhou},
  \citenamefont {Xia}, \citenamefont {Luo}, \citenamefont {Huang},
  \citenamefont {Zhang}, \citenamefont {Yan}, \citenamefont {Fan},
  \citenamefont {Wu}, \citenamefont {Chen}, \citenamefont {Lu},\ and\
  \citenamefont {Hu}}]{Exp_IR_Stability}%
  \BibitemOpen
  \bibfield  {author} {\bibinfo {author} {\bibfnamefont {M.}~\bibnamefont
  {Long}}, \bibinfo {author} {\bibfnamefont {Y.}~\bibnamefont {Wang}}, \bibinfo
  {author} {\bibfnamefont {P.}~\bibnamefont {Wang}}, \bibinfo {author}
  {\bibfnamefont {X.}~\bibnamefont {Zhou}}, \bibinfo {author} {\bibfnamefont
  {H.}~\bibnamefont {Xia}}, \bibinfo {author} {\bibfnamefont {C.}~\bibnamefont
  {Luo}}, \bibinfo {author} {\bibfnamefont {S.}~\bibnamefont {Huang}}, \bibinfo
  {author} {\bibfnamefont {G.}~\bibnamefont {Zhang}}, \bibinfo {author}
  {\bibfnamefont {H.}~\bibnamefont {Yan}}, \bibinfo {author} {\bibfnamefont
  {Z.}~\bibnamefont {Fan}}, \bibinfo {author} {\bibfnamefont {X.}~\bibnamefont
  {Wu}}, \bibinfo {author} {\bibfnamefont {X.}~\bibnamefont {Chen}}, \bibinfo
  {author} {\bibfnamefont {W.}~\bibnamefont {Lu}}, \ and\ \bibinfo {author}
  {\bibfnamefont {W.}~\bibnamefont {Hu}},\ }\href@noop {} {\bibfield  {journal}
  {\bibinfo  {journal} {ACS Nano}\ }\textbf {\bibinfo {volume} {13}},\ \bibinfo
  {pages} {2511} (\bibinfo {year} {2019})}\BibitemShut {NoStop}%
\bibitem [{\citenamefont {Wang}\ \emph {et~al.}(2022)\citenamefont {Wang},
  \citenamefont {Zhou}, \citenamefont {Zhang}, \citenamefont {Xiao},
  \citenamefont {Xu},\ and\ \citenamefont {Tsang}}]{IRex1}%
  \BibitemOpen
  \bibfield  {author} {\bibinfo {author} {\bibfnamefont {Y.}~\bibnamefont
  {Wang}}, \bibinfo {author} {\bibfnamefont {Y.}~\bibnamefont {Zhou}}, \bibinfo
  {author} {\bibfnamefont {Z.}~\bibnamefont {Zhang}}, \bibinfo {author}
  {\bibfnamefont {S.}~\bibnamefont {Xiao}}, \bibinfo {author} {\bibfnamefont
  {J.-b.}\ \bibnamefont {Xu}}, \ and\ \bibinfo {author} {\bibfnamefont {H.~K.}\
  \bibnamefont {Tsang}},\ }\href@noop {} {\bibfield  {journal} {\bibinfo
  {journal} {Appl. Phys. Lett.}\ }\textbf {\bibinfo {volume} {120}},\ \bibinfo
  {pages} {231102} (\bibinfo {year} {2022})}\BibitemShut {NoStop}%
\bibitem [{\citenamefont {Sefidmooye~Azar}\ \emph {et~al.}(2021)\citenamefont
  {Sefidmooye~Azar}, \citenamefont {Bullock}, \citenamefont {Shrestha},
  \citenamefont {Balendhran}, \citenamefont {Yan}, \citenamefont {Kim},
  \citenamefont {Javey},\ and\ \citenamefont {Crozier}}]{IRex2}%
  \BibitemOpen
  \bibfield  {author} {\bibinfo {author} {\bibfnamefont {N.}~\bibnamefont
  {Sefidmooye~Azar}}, \bibinfo {author} {\bibfnamefont {J.}~\bibnamefont
  {Bullock}}, \bibinfo {author} {\bibfnamefont {V.~R.}\ \bibnamefont
  {Shrestha}}, \bibinfo {author} {\bibfnamefont {S.}~\bibnamefont
  {Balendhran}}, \bibinfo {author} {\bibfnamefont {W.}~\bibnamefont {Yan}},
  \bibinfo {author} {\bibfnamefont {H.}~\bibnamefont {Kim}}, \bibinfo {author}
  {\bibfnamefont {A.}~\bibnamefont {Javey}}, \ and\ \bibinfo {author}
  {\bibfnamefont {K.~B.}\ \bibnamefont {Crozier}},\ }\href@noop {} {\bibfield
  {journal} {\bibinfo  {journal} {ACS Nano}\ }\textbf {\bibinfo {volume}
  {15}},\ \bibinfo {pages} {6573} (\bibinfo {year} {2021})}\BibitemShut
  {NoStop}%
\bibitem [{\citenamefont {Island}\ \emph {et~al.}(2015)\citenamefont {Island},
  \citenamefont {Steele}, \citenamefont {van~der Zant},\ and\ \citenamefont
  {Castellanos-Gomez}}]{Island2015}%
  \BibitemOpen
  \bibfield  {author} {\bibinfo {author} {\bibfnamefont {J.~O.}\ \bibnamefont
  {Island}}, \bibinfo {author} {\bibfnamefont {G.~A.}\ \bibnamefont {Steele}},
  \bibinfo {author} {\bibfnamefont {H.~S.}\ \bibnamefont {van~der Zant}}, \
  and\ \bibinfo {author} {\bibfnamefont {A.}~\bibnamefont
  {Castellanos-Gomez}},\ }\href@noop {} {\bibfield  {journal} {\bibinfo
  {journal} {2D Mater.}\ }\textbf {\bibinfo {volume} {2}},\ \bibinfo {pages}
  {011002} (\bibinfo {year} {2015})}\BibitemShut {NoStop}%
\bibitem [{\citenamefont {Artel}\ \emph {et~al.}(2017)\citenamefont {Artel},
  \citenamefont {Guo}, \citenamefont {Cohen}, \citenamefont {Liao},
  \citenamefont {Yang}, \citenamefont {Zhang}, \citenamefont {Lin},
  \citenamefont {Chen}, \citenamefont {Zhou}, \citenamefont {Liu},
  \citenamefont {Shi}, \citenamefont {Wu}, \citenamefont {Zhang},\ and\
  \citenamefont {Wang}}]{Artel2017}%
  \BibitemOpen
  \bibfield  {author} {\bibinfo {author} {\bibfnamefont {V.}~\bibnamefont
  {Artel}}, \bibinfo {author} {\bibfnamefont {Q.}~\bibnamefont {Guo}}, \bibinfo
  {author} {\bibfnamefont {H.}~\bibnamefont {Cohen}}, \bibinfo {author}
  {\bibfnamefont {Y.}~\bibnamefont {Liao}}, \bibinfo {author} {\bibfnamefont
  {S.}~\bibnamefont {Yang}}, \bibinfo {author} {\bibfnamefont {W.}~\bibnamefont
  {Zhang}}, \bibinfo {author} {\bibfnamefont {J.}~\bibnamefont {Lin}}, \bibinfo
  {author} {\bibfnamefont {X.}~\bibnamefont {Chen}}, \bibinfo {author}
  {\bibfnamefont {G.}~\bibnamefont {Zhou}}, \bibinfo {author} {\bibfnamefont
  {X.}~\bibnamefont {Liu}}, \bibinfo {author} {\bibfnamefont {Y.}~\bibnamefont
  {Shi}}, \bibinfo {author} {\bibfnamefont {Y.}~\bibnamefont {Wu}}, \bibinfo
  {author} {\bibfnamefont {Z.}~\bibnamefont {Zhang}}, \ and\ \bibinfo {author}
  {\bibfnamefont {X.}~\bibnamefont {Wang}},\ }\href@noop {} {\bibfield
  {journal} {\bibinfo  {journal} {npj 2D Materials and Applications}\ }\textbf
  {\bibinfo {volume} {1}},\ \bibinfo {pages} {6} (\bibinfo {year}
  {2017})}\BibitemShut {NoStop}%
\bibitem [{\citenamefont {Debnath}, \citenamefont {Park},\ and\ \citenamefont
  {Song}(2018)}]{Review_bP}%
  \BibitemOpen
  \bibfield  {author} {\bibinfo {author} {\bibfnamefont {P.~C.}\ \bibnamefont
  {Debnath}}, \bibinfo {author} {\bibfnamefont {K.}~\bibnamefont {Park}}, \
  and\ \bibinfo {author} {\bibfnamefont {Y.-W.}\ \bibnamefont {Song}},\
  }\href@noop {} {\bibfield  {journal} {\bibinfo  {journal} {Small Methods}\
  }\textbf {\bibinfo {volume} {2}},\ \bibinfo {pages} {1700315} (\bibinfo
  {year} {2018})}\BibitemShut {NoStop}%
\bibitem [{\citenamefont {Dong}\ \emph {et~al.}(2021)\citenamefont {Dong},
  \citenamefont {Yu}, \citenamefont {Zhang}, \citenamefont {Mu}, \citenamefont
  {Xie}, \citenamefont {Li}, \citenamefont {Zhang}, \citenamefont {Huang},
  \citenamefont {He}, \citenamefont {Wang}, \citenamefont {Lin},\ and\
  \citenamefont {Zhang}}]{BBex1}%
  \BibitemOpen
  \bibfield  {author} {\bibinfo {author} {\bibfnamefont {Z.}~\bibnamefont
  {Dong}}, \bibinfo {author} {\bibfnamefont {W.}~\bibnamefont {Yu}}, \bibinfo
  {author} {\bibfnamefont {L.}~\bibnamefont {Zhang}}, \bibinfo {author}
  {\bibfnamefont {H.}~\bibnamefont {Mu}}, \bibinfo {author} {\bibfnamefont
  {L.}~\bibnamefont {Xie}}, \bibinfo {author} {\bibfnamefont {J.}~\bibnamefont
  {Li}}, \bibinfo {author} {\bibfnamefont {Y.}~\bibnamefont {Zhang}}, \bibinfo
  {author} {\bibfnamefont {L.}~\bibnamefont {Huang}}, \bibinfo {author}
  {\bibfnamefont {X.}~\bibnamefont {He}}, \bibinfo {author} {\bibfnamefont
  {L.}~\bibnamefont {Wang}}, \bibinfo {author} {\bibfnamefont {S.}~\bibnamefont
  {Lin}}, \ and\ \bibinfo {author} {\bibfnamefont {K.}~\bibnamefont {Zhang}},\
  }\href@noop {} {\bibfield  {journal} {\bibinfo  {journal} {ACS Nano}\
  }\textbf {\bibinfo {volume} {15}},\ \bibinfo {pages} {20403} (\bibinfo {year}
  {2021})}\BibitemShut {NoStop}%
\bibitem [{\citenamefont {Zeng}\ \emph {et~al.}(2019)\citenamefont {Zeng},
  \citenamefont {Wu}, \citenamefont {Lin}, \citenamefont {Xie}, \citenamefont
  {Yuan}, \citenamefont {Lu}, \citenamefont {Lau}, \citenamefont {Chai},
  \citenamefont {Luo}, \citenamefont {Li},\ and\ \citenamefont
  {Tsang}}]{Exp2_Abs_BBex2}%
  \BibitemOpen
  \bibfield  {author} {\bibinfo {author} {\bibfnamefont {L.-H.}\ \bibnamefont
  {Zeng}}, \bibinfo {author} {\bibfnamefont {D.}~\bibnamefont {Wu}}, \bibinfo
  {author} {\bibfnamefont {S.-H.}\ \bibnamefont {Lin}}, \bibinfo {author}
  {\bibfnamefont {C.}~\bibnamefont {Xie}}, \bibinfo {author} {\bibfnamefont
  {H.-Y.}\ \bibnamefont {Yuan}}, \bibinfo {author} {\bibfnamefont
  {W.}~\bibnamefont {Lu}}, \bibinfo {author} {\bibfnamefont {S.~P.}\
  \bibnamefont {Lau}}, \bibinfo {author} {\bibfnamefont {Y.}~\bibnamefont
  {Chai}}, \bibinfo {author} {\bibfnamefont {L.-B.}\ \bibnamefont {Luo}},
  \bibinfo {author} {\bibfnamefont {Z.-J.}\ \bibnamefont {Li}}, \ and\ \bibinfo
  {author} {\bibfnamefont {Y.~H.}\ \bibnamefont {Tsang}},\ }\href@noop {}
  {\bibfield  {journal} {\bibinfo  {journal} {Adv. Functional Mater.}\ }\textbf
  {\bibinfo {volume} {29}},\ \bibinfo {pages} {1806878} (\bibinfo {year}
  {2019})}\BibitemShut {NoStop}%
\bibitem [{\citenamefont {Wu}\ \emph {et~al.}(2019)\citenamefont {Wu},
  \citenamefont {Guo}, \citenamefont {Du}, \citenamefont {Xia}, \citenamefont
  {Zeng}, \citenamefont {Tian}, \citenamefont {Shi}, \citenamefont {Tian},
  \citenamefont {Li}, \citenamefont {Tsang},\ and\ \citenamefont
  {Jie}}]{BBex3}%
  \BibitemOpen
  \bibfield  {author} {\bibinfo {author} {\bibfnamefont {D.}~\bibnamefont
  {Wu}}, \bibinfo {author} {\bibfnamefont {J.}~\bibnamefont {Guo}}, \bibinfo
  {author} {\bibfnamefont {J.}~\bibnamefont {Du}}, \bibinfo {author}
  {\bibfnamefont {C.}~\bibnamefont {Xia}}, \bibinfo {author} {\bibfnamefont
  {L.}~\bibnamefont {Zeng}}, \bibinfo {author} {\bibfnamefont {Y.}~\bibnamefont
  {Tian}}, \bibinfo {author} {\bibfnamefont {Z.}~\bibnamefont {Shi}}, \bibinfo
  {author} {\bibfnamefont {Y.}~\bibnamefont {Tian}}, \bibinfo {author}
  {\bibfnamefont {X.~J.}\ \bibnamefont {Li}}, \bibinfo {author} {\bibfnamefont
  {Y.~H.}\ \bibnamefont {Tsang}}, \ and\ \bibinfo {author} {\bibfnamefont
  {J.}~\bibnamefont {Jie}},\ }\href@noop {} {\bibfield  {journal} {\bibinfo
  {journal} {ACS Nano}\ }\textbf {\bibinfo {volume} {13}},\ \bibinfo {pages}
  {9907} (\bibinfo {year} {2019})}\BibitemShut {NoStop}%
\bibitem [{\citenamefont {Chen}\ \emph {et~al.}(2021)\citenamefont {Chen},
  \citenamefont {Huang}, \citenamefont {Chen}, \citenamefont {Chen},
  \citenamefont {Hu}, \citenamefont {Wang},\ and\ \citenamefont
  {Dong}}]{BBex4}%
  \BibitemOpen
  \bibfield  {author} {\bibinfo {author} {\bibfnamefont {X.}~\bibnamefont
  {Chen}}, \bibinfo {author} {\bibfnamefont {J.}~\bibnamefont {Huang}},
  \bibinfo {author} {\bibfnamefont {C.}~\bibnamefont {Chen}}, \bibinfo {author}
  {\bibfnamefont {M.}~\bibnamefont {Chen}}, \bibinfo {author} {\bibfnamefont
  {G.}~\bibnamefont {Hu}}, \bibinfo {author} {\bibfnamefont {H.}~\bibnamefont
  {Wang}}, \ and\ \bibinfo {author} {\bibfnamefont {N.}~\bibnamefont {Dong}},\
  }\href@noop {} {\bibfield  {journal} {\bibinfo  {journal} {Adv. Opt. Mater.}\
  }\textbf {\bibinfo {volume} {10}},\ \bibinfo {pages} {2101963} (\bibinfo
  {year} {2021})}\BibitemShut {NoStop}%
\bibitem [{\citenamefont {Yu}\ \emph {et~al.}(2020)\citenamefont {Yu},
  \citenamefont {Kuang}, \citenamefont {Gao}, \citenamefont {Wang},
  \citenamefont {Chen}, \citenamefont {Ding}, \citenamefont {Liu},
  \citenamefont {Cong}, \citenamefont {He}, \citenamefont {Liu},\ and\
  \citenamefont {Liu}}]{Dichroism}%
  \BibitemOpen
  \bibfield  {author} {\bibinfo {author} {\bibfnamefont {J.}~\bibnamefont
  {Yu}}, \bibinfo {author} {\bibfnamefont {X.}~\bibnamefont {Kuang}}, \bibinfo
  {author} {\bibfnamefont {Y.}~\bibnamefont {Gao}}, \bibinfo {author}
  {\bibfnamefont {Y.}~\bibnamefont {Wang}}, \bibinfo {author} {\bibfnamefont
  {K.}~\bibnamefont {Chen}}, \bibinfo {author} {\bibfnamefont {Z.}~\bibnamefont
  {Ding}}, \bibinfo {author} {\bibfnamefont {J.}~\bibnamefont {Liu}}, \bibinfo
  {author} {\bibfnamefont {C.}~\bibnamefont {Cong}}, \bibinfo {author}
  {\bibfnamefont {J.}~\bibnamefont {He}}, \bibinfo {author} {\bibfnamefont
  {Z.}~\bibnamefont {Liu}}, \ and\ \bibinfo {author} {\bibfnamefont
  {Y.}~\bibnamefont {Liu}},\ }\href@noop {} {\bibfield  {journal} {\bibinfo
  {journal} {Nano Lett.}\ }\textbf {\bibinfo {volume} {20}},\ \bibinfo {pages}
  {1172} (\bibinfo {year} {2020})}\BibitemShut {NoStop}%
\bibitem [{\citenamefont {Liang}\ \emph {et~al.}(2022)\citenamefont {Liang},
  \citenamefont {Chen}, \citenamefont {Zhang},\ and\ \citenamefont
  {Wee}}]{Review_PdSe2}%
  \BibitemOpen
  \bibfield  {author} {\bibinfo {author} {\bibfnamefont {Q.}~\bibnamefont
  {Liang}}, \bibinfo {author} {\bibfnamefont {Z.}~\bibnamefont {Chen}},
  \bibinfo {author} {\bibfnamefont {Q.}~\bibnamefont {Zhang}}, \ and\ \bibinfo
  {author} {\bibfnamefont {A.~T.~S.}\ \bibnamefont {Wee}},\ }\href@noop {}
  {\bibfield  {journal} {\bibinfo  {journal} {Adv. Functional Mater.}\ }\textbf
  {\bibinfo {volume} {32}},\ \bibinfo {pages} {2203555} (\bibinfo {year}
  {2022})}\BibitemShut {NoStop}%
\bibitem [{\citenamefont {Chow}\ \emph {et~al.}(2017)\citenamefont {Chow},
  \citenamefont {Yu}, \citenamefont {Liu}, \citenamefont {Hong}, \citenamefont
  {Wang}, \citenamefont {Zeng}, \citenamefont {Hsu}, \citenamefont {Zhu},
  \citenamefont {Zhou}, \citenamefont {Wang}, \citenamefont {Xia},
  \citenamefont {Yan}, \citenamefont {Chen}, \citenamefont {Wu}, \citenamefont
  {Yu}, \citenamefont {Shen}, \citenamefont {Lin}, \citenamefont {Jin},
  \citenamefont {Tay},\ and\ \citenamefont {Liu}}]{Ambipolar_mobility}%
  \BibitemOpen
  \bibfield  {author} {\bibinfo {author} {\bibfnamefont {W.~L.}\ \bibnamefont
  {Chow}}, \bibinfo {author} {\bibfnamefont {P.}~\bibnamefont {Yu}}, \bibinfo
  {author} {\bibfnamefont {F.}~\bibnamefont {Liu}}, \bibinfo {author}
  {\bibfnamefont {J.}~\bibnamefont {Hong}}, \bibinfo {author} {\bibfnamefont
  {X.}~\bibnamefont {Wang}}, \bibinfo {author} {\bibfnamefont {Q.}~\bibnamefont
  {Zeng}}, \bibinfo {author} {\bibfnamefont {C.-H.}\ \bibnamefont {Hsu}},
  \bibinfo {author} {\bibfnamefont {C.}~\bibnamefont {Zhu}}, \bibinfo {author}
  {\bibfnamefont {J.}~\bibnamefont {Zhou}}, \bibinfo {author} {\bibfnamefont
  {X.}~\bibnamefont {Wang}}, \bibinfo {author} {\bibfnamefont {J.}~\bibnamefont
  {Xia}}, \bibinfo {author} {\bibfnamefont {J.}~\bibnamefont {Yan}}, \bibinfo
  {author} {\bibfnamefont {Y.}~\bibnamefont {Chen}}, \bibinfo {author}
  {\bibfnamefont {D.}~\bibnamefont {Wu}}, \bibinfo {author} {\bibfnamefont
  {T.}~\bibnamefont {Yu}}, \bibinfo {author} {\bibfnamefont {Z.}~\bibnamefont
  {Shen}}, \bibinfo {author} {\bibfnamefont {H.}~\bibnamefont {Lin}}, \bibinfo
  {author} {\bibfnamefont {C.}~\bibnamefont {Jin}}, \bibinfo {author}
  {\bibfnamefont {B.~K.}\ \bibnamefont {Tay}}, \ and\ \bibinfo {author}
  {\bibfnamefont {Z.}~\bibnamefont {Liu}},\ }\href@noop {} {\bibfield
  {journal} {\bibinfo  {journal} {Adv. Mater.}\ }\textbf {\bibinfo {volume}
  {29}},\ \bibinfo {pages} {1602969} (\bibinfo {year} {2017})}\BibitemShut
  {NoStop}%
\bibitem [{\citenamefont {Cattelan}\ \emph {et~al.}(2021)\citenamefont
  {Cattelan}, \citenamefont {Sayers}, \citenamefont {Wolverson},\ and\
  \citenamefont {Carpene}}]{Exp_ARPES}%
  \BibitemOpen
  \bibfield  {author} {\bibinfo {author} {\bibfnamefont {M.}~\bibnamefont
  {Cattelan}}, \bibinfo {author} {\bibfnamefont {C.~J.}\ \bibnamefont
  {Sayers}}, \bibinfo {author} {\bibfnamefont {D.}~\bibnamefont {Wolverson}}, \
  and\ \bibinfo {author} {\bibfnamefont {E.}~\bibnamefont {Carpene}},\
  }\href@noop {} {\bibfield  {journal} {\bibinfo  {journal} {2D Mater.}\
  }\textbf {\bibinfo {volume} {8}},\ \bibinfo {pages} {045036} (\bibinfo {year}
  {2021})}\BibitemShut {NoStop}%
\bibitem [{\citenamefont {Hulliger}(1965)}]{OldEXP}%
  \BibitemOpen
  \bibfield  {author} {\bibinfo {author} {\bibfnamefont {F.}~\bibnamefont
  {Hulliger}},\ }\href@noop {} {\bibfield  {journal} {\bibinfo  {journal} {J.
  Phys. Chem. Solids}\ }\textbf {\bibinfo {volume} {26}},\ \bibinfo {pages}
  {639} (\bibinfo {year} {1965})}\BibitemShut {NoStop}%
\bibitem [{\citenamefont {Nishiyama}\ \emph
  {et~al.}(2022{\natexlab{a}})\citenamefont {Nishiyama}, \citenamefont
  {Nishimura}, \citenamefont {Nishioka}, \citenamefont {Ueno}, \citenamefont
  {Iwamoto},\ and\ \citenamefont {Nagashio}}]{Nishiyama5}%
  \BibitemOpen
  \bibfield  {author} {\bibinfo {author} {\bibfnamefont {W.}~\bibnamefont
  {Nishiyama}}, \bibinfo {author} {\bibfnamefont {T.}~\bibnamefont
  {Nishimura}}, \bibinfo {author} {\bibfnamefont {M.}~\bibnamefont {Nishioka}},
  \bibinfo {author} {\bibfnamefont {K.}~\bibnamefont {Ueno}}, \bibinfo {author}
  {\bibfnamefont {S.}~\bibnamefont {Iwamoto}}, \ and\ \bibinfo {author}
  {\bibfnamefont {K.}~\bibnamefont {Nagashio}},\ }\href@noop {} {\bibfield
  {journal} {\bibinfo  {journal} {Adv, Photon. Res.}\ }\textbf {\bibinfo
  {volume} {3}},\ \bibinfo {pages} {2200231} (\bibinfo {year}
  {2022}{\natexlab{a}})}\BibitemShut {NoStop}%
\bibitem [{\citenamefont {Sun}\ \emph {et~al.}(2015)\citenamefont {Sun},
  \citenamefont {Shi}, \citenamefont {Siegrist},\ and\ \citenamefont
  {Singh}}]{mBJ_Seebeck}%
  \BibitemOpen
  \bibfield  {author} {\bibinfo {author} {\bibfnamefont {J.}~\bibnamefont
  {Sun}}, \bibinfo {author} {\bibfnamefont {H.}~\bibnamefont {Shi}}, \bibinfo
  {author} {\bibfnamefont {T.}~\bibnamefont {Siegrist}}, \ and\ \bibinfo
  {author} {\bibfnamefont {D.~J.}\ \bibnamefont {Singh}},\ }\href@noop {}
  {\bibfield  {journal} {\bibinfo  {journal} {Appl. Phys. Lett.}\ }\textbf
  {\bibinfo {volume} {107}},\ \bibinfo {pages} {153902} (\bibinfo {year}
  {2015})}\BibitemShut {NoStop}%
\bibitem [{\citenamefont {Kim}\ and\ \citenamefont
  {Choi}(2021)}]{Kim_Choi_GWgap}%
  \BibitemOpen
  \bibfield  {author} {\bibinfo {author} {\bibfnamefont {H.-G.}\ \bibnamefont
  {Kim}}\ and\ \bibinfo {author} {\bibfnamefont {H.~J.}\ \bibnamefont {Choi}},\
  }\href@noop {} {\bibfield  {journal} {\bibinfo  {journal} {Phys. Rev. B}\
  }\textbf {\bibinfo {volume} {103}},\ \bibinfo {pages} {165419} (\bibinfo
  {year} {2021})}\BibitemShut {NoStop}%
\bibitem [{\citenamefont {Hybertsen}\ and\ \citenamefont
  {Louie}(1986)}]{GW_Louie}%
  \BibitemOpen
  \bibfield  {author} {\bibinfo {author} {\bibfnamefont {M.~S.}\ \bibnamefont
  {Hybertsen}}\ and\ \bibinfo {author} {\bibfnamefont {S.~G.}\ \bibnamefont
  {Louie}},\ }\href@noop {} {\bibfield  {journal} {\bibinfo  {journal} {Phys.
  Rev. B}\ }\textbf {\bibinfo {volume} {34}},\ \bibinfo {pages} {5390}
  (\bibinfo {year} {1986})}\BibitemShut {NoStop}%
\bibitem [{\citenamefont {Perdew}, \citenamefont {Burke},\ and\ \citenamefont
  {Ernzerhof}(1996)}]{PBE}%
  \BibitemOpen
  \bibfield  {author} {\bibinfo {author} {\bibfnamefont {J.~P.}\ \bibnamefont
  {Perdew}}, \bibinfo {author} {\bibfnamefont {K.}~\bibnamefont {Burke}}, \
  and\ \bibinfo {author} {\bibfnamefont {M.}~\bibnamefont {Ernzerhof}},\
  }\href@noop {} {\bibfield  {journal} {\bibinfo  {journal} {Phys. Rev. Lett.}\
  }\textbf {\bibinfo {volume} {77}},\ \bibinfo {pages} {3865} (\bibinfo {year}
  {1996})}\BibitemShut {NoStop}%
\bibitem [{\citenamefont {Salpeter}\ and\ \citenamefont {Bethe}(1951)}]{BSE}%
  \BibitemOpen
  \bibfield  {author} {\bibinfo {author} {\bibfnamefont {E.~E.}\ \bibnamefont
  {Salpeter}}\ and\ \bibinfo {author} {\bibfnamefont {H.~A.}\ \bibnamefont
  {Bethe}},\ }\href@noop {} {\bibfield  {journal} {\bibinfo  {journal} {Phys.
  Rev.}\ }\textbf {\bibinfo {volume} {84}},\ \bibinfo {pages} {1232} (\bibinfo
  {year} {1951})}\BibitemShut {NoStop}%
\bibitem [{\citenamefont {Deilmann}\ and\ \citenamefont
  {Thygesen}(2019)}]{FinMom1}%
  \BibitemOpen
  \bibfield  {author} {\bibinfo {author} {\bibfnamefont {T.}~\bibnamefont
  {Deilmann}}\ and\ \bibinfo {author} {\bibfnamefont {K.~S.}\ \bibnamefont
  {Thygesen}},\ }\href@noop {} {\bibfield  {journal} {\bibinfo  {journal} {2D
  Mater.}\ }\textbf {\bibinfo {volume} {6}},\ \bibinfo {pages} {035003}
  (\bibinfo {year} {2019})}\BibitemShut {NoStop}%
\bibitem [{\citenamefont {Mei}\ \emph {et~al.}(2022)\citenamefont {Mei},
  \citenamefont {Xia}, \citenamefont {Zhang}, \citenamefont {Wu}, \citenamefont
  {Chen}, \citenamefont {Ma}, \citenamefont {Kong}, \citenamefont {Peng},
  \citenamefont {Zhu},\ and\ \citenamefont {Zhang}}]{FinMom2}%
  \BibitemOpen
  \bibfield  {author} {\bibinfo {author} {\bibfnamefont {H.}~\bibnamefont
  {Mei}}, \bibinfo {author} {\bibfnamefont {Y.}~\bibnamefont {Xia}}, \bibinfo
  {author} {\bibfnamefont {Y.}~\bibnamefont {Zhang}}, \bibinfo {author}
  {\bibfnamefont {Y.}~\bibnamefont {Wu}}, \bibinfo {author} {\bibfnamefont
  {Y.}~\bibnamefont {Chen}}, \bibinfo {author} {\bibfnamefont {C.}~\bibnamefont
  {Ma}}, \bibinfo {author} {\bibfnamefont {M.}~\bibnamefont {Kong}}, \bibinfo
  {author} {\bibfnamefont {L.}~\bibnamefont {Peng}}, \bibinfo {author}
  {\bibfnamefont {H.}~\bibnamefont {Zhu}}, \ and\ \bibinfo {author}
  {\bibfnamefont {H.}~\bibnamefont {Zhang}},\ }\href@noop {} {\bibfield
  {journal} {\bibinfo  {journal} {Phys. Chem. Chem. Phys.}\ }\textbf {\bibinfo
  {volume} {24}},\ \bibinfo {pages} {9384} (\bibinfo {year}
  {2022})}\BibitemShut {NoStop}%
\bibitem [{\citenamefont {Lettmann}\ and\ \citenamefont
  {Rohlfing}(2021)}]{FinMom3}%
  \BibitemOpen
  \bibfield  {author} {\bibinfo {author} {\bibfnamefont {T.}~\bibnamefont
  {Lettmann}}\ and\ \bibinfo {author} {\bibfnamefont {M.}~\bibnamefont
  {Rohlfing}},\ }\href@noop {} {\bibfield  {journal} {\bibinfo  {journal}
  {Phys. Rev. B}\ }\textbf {\bibinfo {volume} {104}},\ \bibinfo {pages}
  {115427} (\bibinfo {year} {2021})}\BibitemShut {NoStop}%
\bibitem [{\citenamefont {Kuklin}\ and\ \citenamefont
  {\AA{}gren}(2019)}]{Kuklin_Agren}%
  \BibitemOpen
  \bibfield  {author} {\bibinfo {author} {\bibfnamefont {A.~V.}\ \bibnamefont
  {Kuklin}}\ and\ \bibinfo {author} {\bibfnamefont {H.}~\bibnamefont
  {\AA{}gren}},\ }\href@noop {} {\bibfield  {journal} {\bibinfo  {journal}
  {Phys. Rev. B}\ }\textbf {\bibinfo {volume} {99}},\ \bibinfo {pages} {245114}
  (\bibinfo {year} {2019})}\BibitemShut {NoStop}%
\bibitem [{\citenamefont {Wing}\ \emph {et~al.}(2019)\citenamefont {Wing},
  \citenamefont {Haber}, \citenamefont {Noff}, \citenamefont {Barker},
  \citenamefont {Egger}, \citenamefont {Ramasubramaniam}, \citenamefont
  {Louie}, \citenamefont {Neaton},\ and\ \citenamefont
  {Kronik}}]{SRSH1_alpBetEps2}%
  \BibitemOpen
  \bibfield  {author} {\bibinfo {author} {\bibfnamefont {D.}~\bibnamefont
  {Wing}}, \bibinfo {author} {\bibfnamefont {J.~B.}\ \bibnamefont {Haber}},
  \bibinfo {author} {\bibfnamefont {R.}~\bibnamefont {Noff}}, \bibinfo {author}
  {\bibfnamefont {B.}~\bibnamefont {Barker}}, \bibinfo {author} {\bibfnamefont
  {D.~A.}\ \bibnamefont {Egger}}, \bibinfo {author} {\bibfnamefont
  {A.}~\bibnamefont {Ramasubramaniam}}, \bibinfo {author} {\bibfnamefont
  {S.~G.}\ \bibnamefont {Louie}}, \bibinfo {author} {\bibfnamefont {J.~B.}\
  \bibnamefont {Neaton}}, \ and\ \bibinfo {author} {\bibfnamefont
  {L.}~\bibnamefont {Kronik}},\ }\href@noop {} {\bibfield  {journal} {\bibinfo
  {journal} {Phys. Rev. Mater.}\ }\textbf {\bibinfo {volume} {3}},\ \bibinfo
  {pages} {064603} (\bibinfo {year} {2019})}\BibitemShut {NoStop}%
\bibitem [{\citenamefont {Ramasubramaniam}, \citenamefont {Wing},\ and\
  \citenamefont {Kronik}(2019)}]{SRSH2}%
  \BibitemOpen
  \bibfield  {author} {\bibinfo {author} {\bibfnamefont {A.}~\bibnamefont
  {Ramasubramaniam}}, \bibinfo {author} {\bibfnamefont {D.}~\bibnamefont
  {Wing}}, \ and\ \bibinfo {author} {\bibfnamefont {L.}~\bibnamefont
  {Kronik}},\ }\href@noop {} {\bibfield  {journal} {\bibinfo  {journal} {Phys.
  Rev. Mater.}\ }\textbf {\bibinfo {volume} {3}},\ \bibinfo {pages} {084007}
  (\bibinfo {year} {2019})}\BibitemShut {NoStop}%
\bibitem [{\citenamefont {Refaely-Abramson}\ \emph {et~al.}(2015)\citenamefont
  {Refaely-Abramson}, \citenamefont {Jain}, \citenamefont {Sharifzadeh},
  \citenamefont {Neaton},\ and\ \citenamefont {Kronik}}]{Refaely-Abramson_15}%
  \BibitemOpen
  \bibfield  {author} {\bibinfo {author} {\bibfnamefont {S.}~\bibnamefont
  {Refaely-Abramson}}, \bibinfo {author} {\bibfnamefont {M.}~\bibnamefont
  {Jain}}, \bibinfo {author} {\bibfnamefont {S.}~\bibnamefont {Sharifzadeh}},
  \bibinfo {author} {\bibfnamefont {J.~B.}\ \bibnamefont {Neaton}}, \ and\
  \bibinfo {author} {\bibfnamefont {L.}~\bibnamefont {Kronik}},\ }\href@noop {}
  {\bibfield  {journal} {\bibinfo  {journal} {Phys. Rev. B}\ }\textbf {\bibinfo
  {volume} {92}},\ \bibinfo {pages} {081204} (\bibinfo {year}
  {2015})}\BibitemShut {NoStop}%
\bibitem [{\citenamefont {Camarasa-G\'omez}\ \emph {et~al.}(2023)\citenamefont
  {Camarasa-G\'omez}, \citenamefont {Ramasubramaniam}, \citenamefont {Neaton},\
  and\ \citenamefont {Kronik}}]{Maria_WOT2}%
  \BibitemOpen
  \bibfield  {author} {\bibinfo {author} {\bibfnamefont {M.}~\bibnamefont
  {Camarasa-G\'omez}}, \bibinfo {author} {\bibfnamefont {A.}~\bibnamefont
  {Ramasubramaniam}}, \bibinfo {author} {\bibfnamefont {J.~B.}\ \bibnamefont
  {Neaton}}, \ and\ \bibinfo {author} {\bibfnamefont {L.}~\bibnamefont
  {Kronik}},\ }\href@noop {} {\bibfield  {journal} {\bibinfo  {journal} {Phys.
  Rev. Mater.}\ }\textbf {\bibinfo {volume} {7}},\ \bibinfo {pages} {104001}
  (\bibinfo {year} {2023})}\BibitemShut {NoStop}%
\bibitem [{\citenamefont {Mostofi}\ \emph {et~al.}(2008)\citenamefont
  {Mostofi}, \citenamefont {Yates}, \citenamefont {Lee}, \citenamefont {Souza},
  \citenamefont {Vanderbilt},\ and\ \citenamefont {Marzari}}]{Wannier90}%
  \BibitemOpen
  \bibfield  {author} {\bibinfo {author} {\bibfnamefont {A.~A.}\ \bibnamefont
  {Mostofi}}, \bibinfo {author} {\bibfnamefont {J.~R.}\ \bibnamefont {Yates}},
  \bibinfo {author} {\bibfnamefont {Y.-S.}\ \bibnamefont {Lee}}, \bibinfo
  {author} {\bibfnamefont {I.}~\bibnamefont {Souza}}, \bibinfo {author}
  {\bibfnamefont {D.}~\bibnamefont {Vanderbilt}}, \ and\ \bibinfo {author}
  {\bibfnamefont {N.}~\bibnamefont {Marzari}},\ }\href@noop {} {\bibfield
  {journal} {\bibinfo  {journal} {Comp. Phys. Commun.}\ }\textbf {\bibinfo
  {volume} {178}},\ \bibinfo {pages} {685} (\bibinfo {year}
  {2008})}\BibitemShut {NoStop}%
\bibitem [{\citenamefont {Wing}\ \emph {et~al.}(2021)\citenamefont {Wing},
  \citenamefont {Ohad}, \citenamefont {Haber}, \citenamefont {Filip},
  \citenamefont {Gant}, \citenamefont {Neaton},\ and\ \citenamefont
  {Kronik}}]{WOT1}%
  \BibitemOpen
  \bibfield  {author} {\bibinfo {author} {\bibfnamefont {D.}~\bibnamefont
  {Wing}}, \bibinfo {author} {\bibfnamefont {G.}~\bibnamefont {Ohad}}, \bibinfo
  {author} {\bibfnamefont {J.~B.}\ \bibnamefont {Haber}}, \bibinfo {author}
  {\bibfnamefont {M.~R.}\ \bibnamefont {Filip}}, \bibinfo {author}
  {\bibfnamefont {S.~E.}\ \bibnamefont {Gant}}, \bibinfo {author}
  {\bibfnamefont {J.~B.}\ \bibnamefont {Neaton}}, \ and\ \bibinfo {author}
  {\bibfnamefont {L.}~\bibnamefont {Kronik}},\ }\href@noop {} {\bibfield
  {journal} {\bibinfo  {journal} {Proc. Natl. Acad. Sci.}\ }\textbf {\bibinfo
  {volume} {118}},\ \bibinfo {pages} {e2104556118} (\bibinfo {year}
  {2021})}\BibitemShut {NoStop}%
\bibitem [{\citenamefont {Ohad}\ \emph {et~al.}(2022)\citenamefont {Ohad},
  \citenamefont {Wing}, \citenamefont {Gant}, \citenamefont {Cohen},
  \citenamefont {Haber}, \citenamefont {Sagredo}, \citenamefont {Filip},
  \citenamefont {Neaton},\ and\ \citenamefont {Kronik}}]{WOT2}%
  \BibitemOpen
  \bibfield  {author} {\bibinfo {author} {\bibfnamefont {G.}~\bibnamefont
  {Ohad}}, \bibinfo {author} {\bibfnamefont {D.}~\bibnamefont {Wing}}, \bibinfo
  {author} {\bibfnamefont {S.~E.}\ \bibnamefont {Gant}}, \bibinfo {author}
  {\bibfnamefont {A.~V.}\ \bibnamefont {Cohen}}, \bibinfo {author}
  {\bibfnamefont {J.~B.}\ \bibnamefont {Haber}}, \bibinfo {author}
  {\bibfnamefont {F.}~\bibnamefont {Sagredo}}, \bibinfo {author} {\bibfnamefont
  {M.~R.}\ \bibnamefont {Filip}}, \bibinfo {author} {\bibfnamefont {J.~B.}\
  \bibnamefont {Neaton}}, \ and\ \bibinfo {author} {\bibfnamefont
  {L.}~\bibnamefont {Kronik}},\ }\href@noop {} {\bibfield  {journal} {\bibinfo
  {journal} {Phys. Rev. Mater.}\ }\textbf {\bibinfo {volume} {6}},\ \bibinfo
  {pages} {104606} (\bibinfo {year} {2022})}\BibitemShut {NoStop}%
\bibitem [{\citenamefont {Gant}\ \emph {et~al.}(2022)\citenamefont {Gant},
  \citenamefont {Haber}, \citenamefont {Filip}, \citenamefont {Sagredo},
  \citenamefont {Wing}, \citenamefont {Ohad}, \citenamefont {Kronik},\ and\
  \citenamefont {Neaton}}]{WOT3}%
  \BibitemOpen
  \bibfield  {author} {\bibinfo {author} {\bibfnamefont {S.~E.}\ \bibnamefont
  {Gant}}, \bibinfo {author} {\bibfnamefont {J.~B.}\ \bibnamefont {Haber}},
  \bibinfo {author} {\bibfnamefont {M.~R.}\ \bibnamefont {Filip}}, \bibinfo
  {author} {\bibfnamefont {F.}~\bibnamefont {Sagredo}}, \bibinfo {author}
  {\bibfnamefont {D.}~\bibnamefont {Wing}}, \bibinfo {author} {\bibfnamefont
  {G.}~\bibnamefont {Ohad}}, \bibinfo {author} {\bibfnamefont {L.}~\bibnamefont
  {Kronik}}, \ and\ \bibinfo {author} {\bibfnamefont {J.~B.}\ \bibnamefont
  {Neaton}},\ }\href@noop {} {\bibfield  {journal} {\bibinfo  {journal} {Phys.
  Rev. Mater.}\ }\textbf {\bibinfo {volume} {6}},\ \bibinfo {pages} {053802}
  (\bibinfo {year} {2022})}\BibitemShut {NoStop}%
\bibitem [{\citenamefont {Ohad}\ \emph {et~al.}(2023)\citenamefont {Ohad},
  \citenamefont {Gant}, \citenamefont {Wing}, \citenamefont {Haber},
  \citenamefont {Camarasa-G\'omez}, \citenamefont {Sagredo}, \citenamefont
  {Filip}, \citenamefont {Neaton},\ and\ \citenamefont {Kronik}}]{WOT4}%
  \BibitemOpen
  \bibfield  {author} {\bibinfo {author} {\bibfnamefont {G.}~\bibnamefont
  {Ohad}}, \bibinfo {author} {\bibfnamefont {S.~E.}\ \bibnamefont {Gant}},
  \bibinfo {author} {\bibfnamefont {D.}~\bibnamefont {Wing}}, \bibinfo {author}
  {\bibfnamefont {J.~B.}\ \bibnamefont {Haber}}, \bibinfo {author}
  {\bibfnamefont {M.}~\bibnamefont {Camarasa-G\'omez}}, \bibinfo {author}
  {\bibfnamefont {F.}~\bibnamefont {Sagredo}}, \bibinfo {author} {\bibfnamefont
  {M.~R.}\ \bibnamefont {Filip}}, \bibinfo {author} {\bibfnamefont {J.~B.}\
  \bibnamefont {Neaton}}, \ and\ \bibinfo {author} {\bibfnamefont
  {L.}~\bibnamefont {Kronik}},\ }\href@noop {} {\bibfield  {journal} {\bibinfo
  {journal} {Phys. Rev. Mater.}\ }\textbf {\bibinfo {volume} {7}},\ \bibinfo
  {pages} {123803} (\bibinfo {year} {2023})}\BibitemShut {NoStop}%
\bibitem [{\citenamefont {Sagredo}\ \emph {et~al.}(2024)\citenamefont
  {Sagredo}, \citenamefont {Gant}, \citenamefont {Ohad}, \citenamefont {Haber},
  \citenamefont {Filip}, \citenamefont {Kronik},\ and\ \citenamefont
  {Neaton}}]{Sagredo}%
  \BibitemOpen
  \bibfield  {author} {\bibinfo {author} {\bibfnamefont {F.}~\bibnamefont
  {Sagredo}}, \bibinfo {author} {\bibfnamefont {S.~E.}\ \bibnamefont {Gant}},
  \bibinfo {author} {\bibfnamefont {G.}~\bibnamefont {Ohad}}, \bibinfo {author}
  {\bibfnamefont {J.~B.}\ \bibnamefont {Haber}}, \bibinfo {author}
  {\bibfnamefont {M.~R.}\ \bibnamefont {Filip}}, \bibinfo {author}
  {\bibfnamefont {L.}~\bibnamefont {Kronik}}, \ and\ \bibinfo {author}
  {\bibfnamefont {J.~B.}\ \bibnamefont {Neaton}},\ }\href@noop {} {\bibfield
  {journal} {\bibinfo  {journal} {Phys. Rev. Mater.}\ }\textbf {\bibinfo
  {volume} {8}},\ \bibinfo {pages} {105401} (\bibinfo {year}
  {2024})}\BibitemShut {NoStop}%
\bibitem [{\citenamefont {Ohad}\ \emph {et~al.}(2024)\citenamefont {Ohad},
  \citenamefont {Hartstein}, \citenamefont {Gould}, \citenamefont {Neaton},\
  and\ \citenamefont {Kronik}}]{Ohad_JChem_24}%
  \BibitemOpen
  \bibfield  {author} {\bibinfo {author} {\bibfnamefont {G.}~\bibnamefont
  {Ohad}}, \bibinfo {author} {\bibfnamefont {M.}~\bibnamefont {Hartstein}},
  \bibinfo {author} {\bibfnamefont {T.}~\bibnamefont {Gould}}, \bibinfo
  {author} {\bibfnamefont {J.~B.}\ \bibnamefont {Neaton}}, \ and\ \bibinfo
  {author} {\bibfnamefont {L.}~\bibnamefont {Kronik}},\ }\href@noop {}
  {\bibfield  {journal} {\bibinfo  {journal} {J. Chem. Theory Comp.}\ }\textbf
  {\bibinfo {volume} {20}},\ \bibinfo {pages} {7168} (\bibinfo {year}
  {2024})}\BibitemShut {NoStop}%
\bibitem [{\citenamefont {Camarasa-Gómez}\ \emph {et~al.}(2024)\citenamefont
  {Camarasa-Gómez}, \citenamefont {Gant}, \citenamefont {Ohad}, \citenamefont
  {Neaton}, \citenamefont {Ramasubramaniam},\ and\ \citenamefont
  {Kronik}}]{Maria_npj}%
  \BibitemOpen
  \bibfield  {author} {\bibinfo {author} {\bibfnamefont {M.}~\bibnamefont
  {Camarasa-Gómez}}, \bibinfo {author} {\bibfnamefont {S.~E.}\ \bibnamefont
  {Gant}}, \bibinfo {author} {\bibfnamefont {G.}~\bibnamefont {Ohad}}, \bibinfo
  {author} {\bibfnamefont {J.~B.}\ \bibnamefont {Neaton}}, \bibinfo {author}
  {\bibfnamefont {A.}~\bibnamefont {Ramasubramaniam}}, \ and\ \bibinfo {author}
  {\bibfnamefont {L.}~\bibnamefont {Kronik}},\ }\href@noop {} {\bibfield
  {journal} {\bibinfo  {journal} {npj Comp. Mater.}\ }\textbf {\bibinfo
  {volume} {10}},\ \bibinfo {pages} {288} (\bibinfo {year} {2024})}\BibitemShut
  {NoStop}%
\bibitem [{\citenamefont {Leininger}\ \emph {et~al.}(1997)\citenamefont
  {Leininger}, \citenamefont {Stoll}, \citenamefont {Werner},\ and\
  \citenamefont {Savin}}]{Coulomb1}%
  \BibitemOpen
  \bibfield  {author} {\bibinfo {author} {\bibfnamefont {T.}~\bibnamefont
  {Leininger}}, \bibinfo {author} {\bibfnamefont {H.}~\bibnamefont {Stoll}},
  \bibinfo {author} {\bibfnamefont {H.-J.}\ \bibnamefont {Werner}}, \ and\
  \bibinfo {author} {\bibfnamefont {A.}~\bibnamefont {Savin}},\ }\href@noop {}
  {\bibfield  {journal} {\bibinfo  {journal} {Chem. Phys. Let.}\ }\textbf
  {\bibinfo {volume} {275}},\ \bibinfo {pages} {151} (\bibinfo {year}
  {1997})}\BibitemShut {NoStop}%
\bibitem [{\citenamefont {Yanai}, \citenamefont {Tew},\ and\ \citenamefont
  {Handy}(2004)}]{Coulomb2}%
  \BibitemOpen
  \bibfield  {author} {\bibinfo {author} {\bibfnamefont {T.}~\bibnamefont
  {Yanai}}, \bibinfo {author} {\bibfnamefont {D.~P.}\ \bibnamefont {Tew}}, \
  and\ \bibinfo {author} {\bibfnamefont {N.~C.}\ \bibnamefont {Handy}},\
  }\href@noop {} {\bibfield  {journal} {\bibinfo  {journal} {Chem. Phys. Let.}\
  }\textbf {\bibinfo {volume} {393}},\ \bibinfo {pages} {51} (\bibinfo {year}
  {2004})}\BibitemShut {NoStop}%
\bibitem [{\citenamefont {Refaely-Abramson}\ \emph {et~al.}(2013)\citenamefont
  {Refaely-Abramson}, \citenamefont {Sharifzadeh}, \citenamefont {Jain},
  \citenamefont {Baer}, \citenamefont {Neaton},\ and\ \citenamefont
  {Kronik}}]{Refaely-Abramson_13}%
  \BibitemOpen
  \bibfield  {author} {\bibinfo {author} {\bibfnamefont {S.}~\bibnamefont
  {Refaely-Abramson}}, \bibinfo {author} {\bibfnamefont {S.}~\bibnamefont
  {Sharifzadeh}}, \bibinfo {author} {\bibfnamefont {M.}~\bibnamefont {Jain}},
  \bibinfo {author} {\bibfnamefont {R.}~\bibnamefont {Baer}}, \bibinfo {author}
  {\bibfnamefont {J.~B.}\ \bibnamefont {Neaton}}, \ and\ \bibinfo {author}
  {\bibfnamefont {L.}~\bibnamefont {Kronik}},\ }\href@noop {} {\bibfield
  {journal} {\bibinfo  {journal} {Phys. Rev. B}\ }\textbf {\bibinfo {volume}
  {88}},\ \bibinfo {pages} {081204} (\bibinfo {year} {2013})}\BibitemShut
  {NoStop}%
\bibitem [{\citenamefont {Cudazzo}, \citenamefont {Tokatly},\ and\
  \citenamefont {Rubio}(2011)}]{DieScreen1}%
  \BibitemOpen
  \bibfield  {author} {\bibinfo {author} {\bibfnamefont {P.}~\bibnamefont
  {Cudazzo}}, \bibinfo {author} {\bibfnamefont {I.~V.}\ \bibnamefont
  {Tokatly}}, \ and\ \bibinfo {author} {\bibfnamefont {A.}~\bibnamefont
  {Rubio}},\ }\href@noop {} {\bibfield  {journal} {\bibinfo  {journal} {Phys.
  Rev. B}\ }\textbf {\bibinfo {volume} {84}},\ \bibinfo {pages} {085406}
  (\bibinfo {year} {2011})}\BibitemShut {NoStop}%
\bibitem [{\citenamefont {Andersen}, \citenamefont {Latini},\ and\
  \citenamefont {Thygesen}(2015)}]{DieScreen2}%
  \BibitemOpen
  \bibfield  {author} {\bibinfo {author} {\bibfnamefont {K.}~\bibnamefont
  {Andersen}}, \bibinfo {author} {\bibfnamefont {S.}~\bibnamefont {Latini}}, \
  and\ \bibinfo {author} {\bibfnamefont {K.~S.}\ \bibnamefont {Thygesen}},\
  }\href@noop {} {\bibfield  {journal} {\bibinfo  {journal} {Nano Let.}\
  }\textbf {\bibinfo {volume} {15}},\ \bibinfo {pages} {4616} (\bibinfo {year}
  {2015})}\BibitemShut {NoStop}%
\bibitem [{\citenamefont {Qiu}, \citenamefont {da~Jornada},\ and\ \citenamefont
  {Louie}(2016)}]{Qiu_etal}%
  \BibitemOpen
  \bibfield  {author} {\bibinfo {author} {\bibfnamefont {D.~Y.}\ \bibnamefont
  {Qiu}}, \bibinfo {author} {\bibfnamefont {F.~H.}\ \bibnamefont {da~Jornada}},
  \ and\ \bibinfo {author} {\bibfnamefont {S.~G.}\ \bibnamefont {Louie}},\
  }\href@noop {} {\bibfield  {journal} {\bibinfo  {journal} {Phys. Rev. B}\
  }\textbf {\bibinfo {volume} {93}},\ \bibinfo {pages} {235435} (\bibinfo
  {year} {2016})}\BibitemShut {NoStop}%
\bibitem [{\citenamefont {Heyd}, \citenamefont {Scuseria},\ and\ \citenamefont
  {Ernzerhof}(2003)}]{HSE06}%
  \BibitemOpen
  \bibfield  {author} {\bibinfo {author} {\bibfnamefont {J.}~\bibnamefont
  {Heyd}}, \bibinfo {author} {\bibfnamefont {G.~E.}\ \bibnamefont {Scuseria}},
  \ and\ \bibinfo {author} {\bibfnamefont {M.}~\bibnamefont {Ernzerhof}},\
  }\href@noop {} {\bibfield  {journal} {\bibinfo  {journal} {J. Chem. Phys.}\
  }\textbf {\bibinfo {volume} {118}},\ \bibinfo {pages} {8207} (\bibinfo {year}
  {2003})}\BibitemShut {NoStop}%
\bibitem [{\citenamefont {Nishiyama}\ \emph
  {et~al.}(2022{\natexlab{b}})\citenamefont {Nishiyama}, \citenamefont
  {Nishimura}, \citenamefont {Ueno}, \citenamefont {Taniguchi}, \citenamefont
  {Watanabe},\ and\ \citenamefont {Nagashio}}]{Nishiyama3}%
  \BibitemOpen
  \bibfield  {author} {\bibinfo {author} {\bibfnamefont {W.}~\bibnamefont
  {Nishiyama}}, \bibinfo {author} {\bibfnamefont {T.}~\bibnamefont
  {Nishimura}}, \bibinfo {author} {\bibfnamefont {K.}~\bibnamefont {Ueno}},
  \bibinfo {author} {\bibfnamefont {T.}~\bibnamefont {Taniguchi}}, \bibinfo
  {author} {\bibfnamefont {K.}~\bibnamefont {Watanabe}}, \ and\ \bibinfo
  {author} {\bibfnamefont {K.}~\bibnamefont {Nagashio}},\ }\href@noop {}
  {\bibfield  {journal} {\bibinfo  {journal} {Adv. Functional Mat.}\ }\textbf
  {\bibinfo {volume} {32}},\ \bibinfo {pages} {2108061} (\bibinfo {year}
  {2022}{\natexlab{b}})}\BibitemShut {NoStop}%
\bibitem [{\citenamefont {Gatti}\ and\ \citenamefont
  {Sottile}(2013)}]{ExcitonDispersion}%
  \BibitemOpen
  \bibfield  {author} {\bibinfo {author} {\bibfnamefont {M.}~\bibnamefont
  {Gatti}}\ and\ \bibinfo {author} {\bibfnamefont {F.}~\bibnamefont
  {Sottile}},\ }\href@noop {} {\bibfield  {journal} {\bibinfo  {journal} {Phys.
  Rev. B}\ }\textbf {\bibinfo {volume} {88}},\ \bibinfo {pages} {155113}
  (\bibinfo {year} {2013})}\BibitemShut {NoStop}%
\bibitem [{\citenamefont {Sander}, \citenamefont {Maggio},\ and\ \citenamefont
  {Kresse}(2015)}]{BeyondTamDan}%
  \BibitemOpen
  \bibfield  {author} {\bibinfo {author} {\bibfnamefont {T.}~\bibnamefont
  {Sander}}, \bibinfo {author} {\bibfnamefont {E.}~\bibnamefont {Maggio}}, \
  and\ \bibinfo {author} {\bibfnamefont {G.}~\bibnamefont {Kresse}},\
  }\href@noop {} {\bibfield  {journal} {\bibinfo  {journal} {Phys. Rev. B}\
  }\textbf {\bibinfo {volume} {92}},\ \bibinfo {pages} {045209} (\bibinfo
  {year} {2015})}\BibitemShut {NoStop}%
\bibitem [{\citenamefont {Qiu}, \citenamefont {Cao},\ and\ \citenamefont
  {Louie}(2015)}]{FinMomExcitons1}%
  \BibitemOpen
  \bibfield  {author} {\bibinfo {author} {\bibfnamefont {D.~Y.}\ \bibnamefont
  {Qiu}}, \bibinfo {author} {\bibfnamefont {T.}~\bibnamefont {Cao}}, \ and\
  \bibinfo {author} {\bibfnamefont {S.~G.}\ \bibnamefont {Louie}},\ }\href@noop
  {} {\bibfield  {journal} {\bibinfo  {journal} {Phys. Rev. Lett.}\ }\textbf
  {\bibinfo {volume} {115}},\ \bibinfo {pages} {176801} (\bibinfo {year}
  {2015})}\BibitemShut {NoStop}%
\bibitem [{\citenamefont {Noffsinger}\ \emph {et~al.}(2012)\citenamefont
  {Noffsinger}, \citenamefont {Kioupakis}, \citenamefont {Van~de Walle},
  \citenamefont {Louie},\ and\ \citenamefont {Cohen}}]{Noffsinger}%
  \BibitemOpen
  \bibfield  {author} {\bibinfo {author} {\bibfnamefont {J.}~\bibnamefont
  {Noffsinger}}, \bibinfo {author} {\bibfnamefont {E.}~\bibnamefont
  {Kioupakis}}, \bibinfo {author} {\bibfnamefont {C.~G.}\ \bibnamefont {Van~de
  Walle}}, \bibinfo {author} {\bibfnamefont {S.~G.}\ \bibnamefont {Louie}}, \
  and\ \bibinfo {author} {\bibfnamefont {M.~L.}\ \bibnamefont {Cohen}},\
  }\href@noop {} {\bibfield  {journal} {\bibinfo  {journal} {Phys. Rev. Lett.}\
  }\textbf {\bibinfo {volume} {108}},\ \bibinfo {pages} {167402} (\bibinfo
  {year} {2012})}\BibitemShut {NoStop}%
\bibitem [{Note1()}]{Note1}%
  \BibitemOpen
  \bibinfo {note} {The center-of-mass momentum vector, ${\protect \bm {Q}}$, is
  denoted here as a multiple of the reciprocal lattice vectors. The band-edge
  extrema obtained from Wannier interpolation (Fig.\protect \,\ref
  {fig:Ebands}) do not necessarily coincide with sampling points in the regular
  $k$-point mesh used for self-consistent calculations. As VASP requires
  $\protect \bm {Q}$ to be expressible as a difference between two $k$-points,
  not allowing for interpolation to irregular points, we pick $k$-points that
  are as close as possible to the valence band maximum and the conduction band
  minimum. Thus, our estimates for optical gaps are upper bounds. The resulting
  overestimate of the optical gap can be approximated by the difference of the
  electronic band gap between the two sampled points and the true indirect
  fundamental band gap, and we estimate this error to be 13.5 meV for the bulk
  and 5.3 meV for the monolayer. The error can be reduced with denser $k$-point
  grids.}\BibitemShut {Stop}%
\bibitem [{\citenamefont {Ramasubramaniam}(2012)}]{AR2012}%
  \BibitemOpen
  \bibfield  {author} {\bibinfo {author} {\bibfnamefont {A.}~\bibnamefont
  {Ramasubramaniam}},\ }\href@noop {} {\bibfield  {journal} {\bibinfo
  {journal} {Phys. Rev. B}\ }\textbf {\bibinfo {volume} {86}},\ \bibinfo
  {pages} {115409} (\bibinfo {year} {2012})}\BibitemShut {NoStop}%
\bibitem [{\citenamefont {Grosso}\ and\ \citenamefont
  {Parravicini}(2000)}]{grosso2000solid}%
  \BibitemOpen
  \bibfield  {author} {\bibinfo {author} {\bibfnamefont {G.}~\bibnamefont
  {Grosso}}\ and\ \bibinfo {author} {\bibfnamefont {G.}~\bibnamefont
  {Parravicini}},\ }\href {https://books.google.com/books?id=L5RrQbbvWn8C}
  {\emph {\bibinfo {title} {Solid State Physics}}}\ (\bibinfo  {publisher}
  {Academic Press},\ \bibinfo {year} {2000})\BibitemShut {NoStop}%
\bibitem [{Note2()}]{Note2}%
  \BibitemOpen
  \bibinfo {note} {Note that the actual absorption spectrum will contain
  contributions from multiple $\protect \bm {Q}$-vectors but we are only
  concerned with one momentum here to capture the absorption
  onset.}\BibitemShut {Stop}%
\end{thebibliography}%

\end{document}